\documentclass[11pt]{article}

\pdfoutput=1

\usepackage{url,alltt}
\usepackage{paralist, xspace}
\usepackage{caption}
\usepackage{subcaption}
\usepackage{graphicx}
\usepackage{wrapfig}
\usepackage{times}
\usepackage{amssymb,amsmath,pifont,footnote}
\usepackage{latexsym,pdfpages}
\usepackage{array,verbatim}
\usepackage{ifsym}

\usepackage{listings}
\usepackage{color}
\usepackage{hyperref}
\usepackage{cleveref}
\usepackage{enumitem}
\usepackage{amsthm}
\usepackage{pdfpages}
\usepackage{bbm}

\usepackage{amsfonts}
\usepackage{fullpage}
\usepackage{float}
\usepackage[vlined,ruled,linesnumbered]{algorithm2e}
\usepackage{thm-restate}

\definecolor{dkgreen}{rgb}{0,0.6,0}
\definecolor{gray}{rgb}{0.5,0.5,0.5}
\definecolor{mauve}{rgb}{0.58,0,0.82}

\setlist{nolistsep}

\newcommand{\BeginProof}{\vspace{-0.25cm}\begin{proof}}

\newcommand{\Comment}[1]{}
\newcommand{\Appendix}[1]{}

\newcommand{\AppendixContent}[1]{}

\newcommand\eat[1]{}

\newcommand{\allnotes}[1]{}
\renewcommand{\allnotes}[1]{\textit{#1}}

\graphicspath{ {./PODC-2018/figures/} }

\newcommand{\Variant}[2]{#2}

\newcommand{\SectionText}{Section }

\newcommand{\sectionText}{section }
\newcommand{\subsomething}{\subsection}
\newcommand{\mySection}{\section}
\newcommand{\SubSomethingText}{Subsection }

\RestyleAlgo{boxruled}
\LinesNumbered

\declaretheorem[name=Theorem,numberwithin=section]{theorem}
\newtheorem{remark}{Remark}
\newtheorem{lemma}[theorem]{Lemma}
\newtheorem{claim}[theorem]{Claim}
\newtheorem{conjecture}[theorem]{Conjecture}
\newtheorem{corollary}[theorem]{Corollary}

\newtheorem{definition}[theorem]{Definition}

\makeatletter
\def\moverlay{\mathpalette\mov@rlay}
\def\mov@rlay#1#2{\leavevmode\vtop{%
        \baselineskip\z@skip \lineskiplimit-\maxdimen
        \ialign{\hfil$\m@th#1##$\hfil\cr#2\crcr}}}
\newcommand{\charfusion}[3][\mathord]{
    #1{\ifx#1\mathop\vphantom{#2}\fi
        \mathpalette\mov@rlay{#2\cr#3}
    }
    \ifx#1\mathop\expandafter\displaylimits\fi}
\makeatother

\newcommand{\cupdot}{\charfusion[\mathbin]{\cup}{\cdot}}

\SetKwProg{Fn}{Function}{}{end}

\newcommand{\plusplus}{\mathrel{{+}{+}}}

\newenvironment{proofoutline}[1]
{\proof[Proof outline for #1]}
{\endproof}

\newcommand*\mean[1]{\bar{#1}}

\newcommand{\RingOriginalProtocol}{{\tt A}-${\tt LEAD}^{\tt uni}$}
\newcommand{\RingOriginalProtocolSpaced}{\RingOriginalProtocol $~$}
\newcommand{\RingOriginalProtocolVerySpaced}{{\tt A}-${\tt LEAD}^{\tt uni}$ $~~$}
\newcommand{\RingOriginalProtocolTitle}{\texorpdfstring{\RingOriginalProtocol}{A-LEAD(uni)}}
\newcommand{\PhaseAsyncLead}{\texttt{PhaseAsyncLead} }
\newcommand{\PhaseAsyncLeadTitle}{\texorpdfstring{\PhaseAsyncLead}{PhaseAsyncLead}}

\newcommand{\kunbiased}[0]{$\epsilon$-$k$-\textit{unbiased }}

\newcommand{\kresilience}[0]{$\epsilon$-$k$-\textit{resilience }}

\newcommand{\kresilient}[0]{$\epsilon$-$k$-\textit{resilient }}
\newcommand{\perfectkresilient}[0]{$k$-\textit{resilient }}

\newcommand{\DataValues}{\{d_{i}\}_{i=1}^{n}}
\newcommand{\ValidationValues}{\{v_{i}\}_{i=1}^{n}}
\newcommand{\SumDataValues}{\sum_{i=1}^{n}{d_{i}} \pmod n}

\newcommand{\FAIL}{{\tt FAIL}}

\begin{document}
	
	\title{
		Fair Leader Election for Rational Agents in Asynchronous Rings and 
		Networks
	}
	
	\author{
		Assaf Yifrach\thanks{Supported in part by a grant from the Israel 
			Science
			Foundation}\\
		Tel-Aviv University\\
		\texttt{asafyi@gmail.com}
		\and
		Yishay Mansour\thanks{Supported in part by a grant from the Israel 
			Science
			Foundation}\\
		Tel-Aviv University \& Google Research\\
		\texttt{mansour.yishay@gmail.com}
	}
	
	\maketitle

\begin{abstract}
 We  study a game theoretic model where a coalition of
 processors might collude to bias the outcome of the protocol, where
 we assume that the processors always prefer any legitimate outcome
 over a non-legitimate one.
 We show that the problems of Fair Leader
 Election and Fair Coin Toss are equivalent, and focus on Fair Leader
 Election.

 Our main focus is on a directed asynchronous ring of $n$ processors,
 where we investigate the protocol proposed by Abraham et al.
 \cite{abraham2013distributed} and studied in Afek et al.
 \cite{afek2014distributed}. We show that in general the protocol is
 resilient only to sub-linear size coalitions. Specifically, we show
 that $\Omega(\sqrt{n\log n})$ randomly located processors or
 $\Omega(\sqrt[3]{n})$ adversarially located processors can force any
 outcome. We complement this by showing that the protocol is
 resilient to any adversarial coalition of size $O(\sqrt[4]{n})$.
 We propose a modification to the protocol, and show that it is
 resilient to every coalition of size $\Theta(\sqrt{n})$, by exhibiting
 both an attack and a resilience result.

 For every $k \geq 1$, we define a family of graphs ${\mathcal{G}}_{k}$
 that can be simulated by trees where each node in the tree simulates
 at most $k$ processors. We show that for every graph in
 ${\mathcal{G}}_{k}$, there is no fair leader election protocol that is
 resilient to coalitions of size $k$. Our result generalizes a
 previous result of Abraham et al. \cite{abraham2013distributed} that
 states that for every graph, there is no fair leader election
 protocol which is resilient to coalitions of size $\lceil \frac{n}{2}
 \rceil$.
\end{abstract}

\mySection{Introduction}
\label{section:Introduction}

One of the most fundamental tasks in distributed computing is fault
tolerance, the ability to overcome malicious or abnormal behavior of
processors. Fault tolerance is essential to make distributed systems
viable, and enables them to operate at large scale and in unsecured
environments.  Different models of fault tolerance assume different
assumptions about faulty processors.  For example, some model assume
that faulty processors are Byzantine, i.e., they can behave in an
arbitrary malicious way. As another example, some models assume that
faulty processors are fail-stop, i.e., they execute the protocol
normally until an arbitrary point and then they stop responding. In
both cases, the only objective of faulty processors is to fail the
protocol.


In this paper, we study protocols that are tolerant to a third type
of faulty processors, \textit{rational agents}. We assume the
processors are selfish. Given a protocol, each processor (which is a
rational agent) has its own utility function over the possible
outcomes of the protocol. A processor deviates from the protocol
(i.e., cheats by running another protocol) if deviating increases
its expected utility.

Later, we explain what reasonable assumptions can be made about the processors' utility functions.

At a high level we would like to design protocols which are
resilient to such deviations. This line of research has been active
for over a decade (see,
\cite{halpern2004rational,moscibroda2006selfish,abraham2006distributed,abraham2011distributed,abraham2013distributed,afek2014distributed,clementi2017rational,afek2017cheating}).

Following this research, we look for solutions (i.e., resilient protocols) in
terms of game-theory. Specifically, we look for a protocol that is a
strong-$k$-Nash-equilibria. That is, a protocol for which there is no
coalition (any subset of the processors) of up to $k$ processors, that can
increase the expected utility of each of its members by deviating
cooperatively. Where deviating cooperatively means running another protocol
instead of the prescribed protocol. As in strong-$k$-Nash-equilibria, such a
coalition assumes that all of the processors outside the coalition play
honestly, i.e., execute the protocol honestly. In other words, two coalitions
cannot deviate in parallel. If a protocol is a strong-$k$-Nash-equilibria for
every set of utility functions, then we  ssay it is $k$-\textit{resilient}.

We explain and motivate our setting using an example. Consider the
problem of designing a leader election protocol for rational agents.
The main issue is that some processors might want to get elected as
a leader in order to gain additional privileges. A natural solution
would be a \textit{Fair} Leader Election protocol, which is a leader
election protocol that elects each processor with equal probability.
One simple protocol, assuming that the $n$ ids of the processors are
$[1,n]$, is to let each processor select a random value in $[n]$ and
broadcast it. Each processor, after receiving all the random values,
can sum the values up (modulus $n$) and the result would be the id
of the elected leader. This simple protocol selects a leader
uniformly, assuming all the processors follow it precisely. However,
in an asynchronous network even a single deviating processor can
control the output and select the leader. A single processor can
cheat by waiting for the random values of all the other processors
to arrive before selecting its own value. Note that this simple
protocol is applicable also in message-passing networks, because one
can implement broadcast over the network.

The weakness of that simple protocol was already observed by Abraham et al.
\cite{abraham2013distributed}, who suggested a methodology to overcome it in a
unidirectional ring network. They named their protocol \RingOriginalProtocol.
The main idea in \RingOriginalProtocolSpaced is to use buffering in order to
delay
the flow of the messages along the ring and thus limit the effect of malicious
processors. They showed that \RingOriginalProtocolSpaced is a
strong-$1$-Nash-equilibria, thus, is overcoming a single malicious processor.
They claimed that their protocol is $(\frac{n}{2}-1)$-resilient, i.e., resilient
to \textbf{every} coalition of size $k \leq \frac{1}{2}n - 1$, however, it is
true only for coalitions that are located \textbf{consecutively} along the ring.

Later, Afek et al. \cite{afek2014distributed} simplified \RingOriginalProtocol, and decomposed it into useful intuitive building blocks.

The main thrust of this paper is studying the resilience of
\RingOriginalProtocolSpaced and improving it.

In our model, we assume the \textit{solution preference} assumption, i.e., that
processors always prefer a selection of any leader over a failure of the
protocol. This assumption is reasonable in various settings. For example,
processors might be able to cheat only during the leader election, which is
usually a preliminary step, but not during the main computation. In such case,
leaving the system in an erroneous state at the termination of the leader
election step, fails also the main computation and thus prevents them from
benefiting from its results.

The solution preference assumption has two benefits. First,  we can
hope for non-trivial resilience results. In a unidirectional ring,
two processors can disconnect the ring and thus fail every
reasonable protocol, however due to the solution preference, a
failure is the worst possible outcome in terms of utility, so they
want to avoid it. Therefore, we can still hope for $k$-resilient
protocols with $k>2$. Second, the solution preference assumption
allows processors to ``punish'' deviating processors. If a processor
detects a deviation, then it aborts the protocol by terminating with
an invalid output and therefore no processor gets elected. Since all
processors know this threatening behavior, a coalition wishes to
bias the output by deviating from the protocol without getting
detected.

In our setting, malicious processors would like to bias the leader election as
much as possible. Our main notion of resilience measures how much the malicious
processors can influence the outcome of the leader election. At a high level,
in our attacks, the malicious processors almost determine the elected
processor. In our resilience results, we prove that the malicious processors
might be able to increase the probability of a processor to get elected only by
a negligible amount.

\bigskip \noindent{\bf Our Contributions:}
 Our primary focus is to find a function
$k=k(n)$ as large as possible, such that there exists a
$k$-resilient leader election protocol for an asynchronous
unidirectional ring. From the other direction, while considering
other topologies of asynchronous networks, we want to find a
function $k=k(n)$ as small as possible, such that there \textbf{does
not exist} a $k$-resilient fair leader election protocol.
In-existence of a $k$-resilient protocol is called an impossibility
claim.

Abraham et al. \cite{abraham2013distributed} presented a uniform leader
election protocol for a unidirectional ring, named \RingOriginalProtocol, and
showed that it is resilient to coalitions of size $k < \frac{1}{2}n$ that are
located \textbf{consecutively} along the ring (for completeness, we also give a
resilience proof of this result in \Variant{the full paper 
\cite{yifrach2018distributed}}{Appendix
\ref{section:ALeadIsResilient}}). For a general asynchronous network, in
particular for a unidirectional ring, Abraham et al. showed that there is no
$k$-resilient protocol for every $k\geq \frac{1}{2}n$.

Our main contributions are:
\begin{itemize}[noitemsep,topsep=0pt]
    \item We give an almost tight resilience analysis for
    \RingOriginalProtocol, for generally located coalitions. First, we show
    that it is \textbf{not} resilient to a randomly located coalition of size
    $k=\Theta(\sqrt{n \log(n)})$ with high probability (\Variant{in the full
    paper}{ Appendix \ref{section:RandomizedModel}}). Then, we also show that
    this protocol is \textbf{not} resilient to $k=2\sqrt[3]{n}$ carefully
    located (i.e., worst case) processors (\SectionText \ref{section:Attacks}).
    Next, we prove that \RingOriginalProtocolSpaced is $k$-resilient for
    $k=O(\sqrt[4]{n})$ (\SectionText \ref{section:GoodBounds}).
    \item We improve \RingOriginalProtocolSpaced by introducing a new protocol \PhaseAsyncLead, a $\Theta(\sqrt{n})$-resilient fair leader election protocol for a unidirectional ring. We exhibit both an attack with $\Theta(\sqrt{n})$ malicious processors, and prove that \PhaseAsyncLead is $k$-resilient for $k=O(\sqrt{n})$.
    \item We generalize a previous impossibility result from \cite{abraham2013distributed}, by showing that there is no $k$-resilient fair leader election protocol for every asynchronous $k$-simulated tree.
    A $k$-simulated tree is a network that can be simulated by a tree network, where each processor in the tree simulates at most $k$ processors. This generalizes the previous impossibility result because any graph is a $\lceil \frac{1}{2}n \rceil$-simulated tree. Also, it strictly improves the previous result because some graphs are $k$-simulated trees for $k \ll \frac{1}{2}n$ (for example, trees are $1$-simulated trees).
    \item Unsurprisingly, we show that Fair Coin Toss and Fair Leader Election 
    are equivalent. Essentially, Fair Coin Toss requires the ability to toss a 
    fair binary coin, while fair leader election requires the ability to toss 
    $\log_2(n)$ binary coins. In order to implement leader election using 
    $log(n)$ coin tosses, we assume the ability to run independent coin tosses. 
    \Variant{We include the details only in the full paper.}{}
\end{itemize}

\bigskip \noindent{\textbf{Our Techniques:}}
The main idea in our attacks on \RingOriginalProtocolSpaced is
rushing the information.  Namely, the attacking processors reduce
the number of messages traversing the ring by not generating their
own random value. This allows them to acquire quickly all the
information that is required to influence the outcome of the
protocol.

The main observation in our resilience proof for
\RingOriginalProtocolSpaced  is that all of the processors must be
``$k^2$-synchronized'' during the execution, or else a deviation is
detected by the honest processors which abort. In this context,
``$m$-synchronized'' means that at every point in time during the
execution, every two processors have sent the same number of
messages up to a difference of $O(m)$.

Another observation used for our resilience proof for
\RingOriginalProtocolSpaced is that the information required for a processor
$p$ in the coalition in order to bias the output is initially located far away.
If the coalition is small enough, then by the time the information reaches $p$,
it is already too late for it to bias the output. This is because $p$ is
committed to what it will send in the future, because the honest processors
validate the contents of all its future messages (honest processors abort if
$p$ does not send the expected messages). For this reason $p$ cannot manipulate
the
output calculated by its honest successor, so in particular the coalition
cannot bias the outcome.

Our main idea in the design of \texttt{PhaseAsyncLead} is forcing
processors  to be more synchronized, specifically, ``$k$ -
synchronized'' instead of ``$k^2$-synchronized''. As a side effect
of the synchronization enforcement in our improved protocol, small
amounts of information might travel quickly, so the technique used
for the previous resilience proof does not apply (as required far
away information can now travel quickly).  In order to cope with
that problem, we use a random function that forces any malicious
processor to obtain a lot of information before being able to bias
the output. We show that due to ``$k$-synchronization'', in order to
get that amount of information, a processor must send a lot of
messages. However, by the time it sends so many messages, it has
already committed to all of its outgoing messages that might affect
the output (i.e., all of its future messages that might affect the
output are validated by other processors as before).

\subsomething{Related Work}
This work, continues the work presented in \cite{abraham2013distributed} by  
Abraham et al. They study resilient protocols for fair leader election in 
message passing networks with rational agents in a variety of scenarios.
\begin{itemize}
	\item For the first two scenarios, a synchronous fully connected network, 
	and a synchronous ring, they suggest optimal solutions which are resilient 
	to $k=n-1$ processors. 
	\item For a scenario with computationally bounded agents under 
	cryptographic assumptions, they provide a similar solution that is based on 
	cryptographic commitments.
	\item For an asynchronous fully connected network, they 
	apply Shamir's secret sharing scheme in a straight-forward manner and get 
	an optimal resilience result of $k=n/2-1$.
	\item For the most complicated scenario, an asynchronous 
	ring, they suggest an interesting protocol and analyze its resilience to 
	only to consecutively located coalitions. However, they do not analyze its 
	resilience to general coalitions. We focus on this scenario, study the 
	resilience of their protocol, and present a more resilient protocol.
\end{itemize}

Additionally, they prove an upper bound for the resilience of fair leader 
election protocols in an asynchronous network. For every asynchronous network, 
there is no fair leader election protocol that is resilient to every coalition 
of size $k=n/2$. We generalize this bound and improve it.

In \cite{afek2014distributed}, Afek et al. re-organize methods suggested in 
\cite{abraham2013distributed} into useful building blocks. Specifically, a 
wake-up building block and a knowledge sharing building block. Additionally, 
they consider protocols for Fair Consensus and for Renaming. Our work, builds 
on their clean reformulation of the protocol suggested by 
\cite{abraham2013distributed}.

Most of our work focuses on the fundamental problem of leader election on a 
directed ring. Standard algorithms for this problem, which are not 
fault tolerant were studied in many classical works, such as
\cite{chang1979improved,dolev1982n,peterson1982n}. 
These classical works, elect the processor with the maximal (or minimal)
id as the leader. They focus on reducing the worst case message complexity and 
the average message complexity of the algorithm. Chang et al. 
\cite{chang1979improved} presented a randomized leader election protocol with 
an average message complexity of  $\Theta(n \log(n)))$, 
while assuming the processors are randomly located along the ring.
Later, Dolev et al. and Peterson et al. \cite{dolev1982n,peterson1982n} 
suggested a deterministic algorithm that improves the worst case message 
complexity to $O(n \log(n))$.

Fault tolerance in distributed systems under classic assumptions of
Byzantine faults and fail-stop faults has been studied extensively.
For examples refer to the following surveys
\cite{correia2011byzantine,sari2015fault}. The survey 
\cite{correia2011byzantine} reviews work on Byzantine consensus in asynchronous 
message passing networks. It presents a few formulations of the problem, and 
points to works that assume different assumptions in order to solve it, such as 
using randomization, or assuming the existence of external failure detectors.

Fault models that combine both Byzantine, and rational processors,
where studied in
\cite{aiyer2005bar,
	abraham2006distributed,abraham2008lower}.
For example, the BAR model suggested in \cite{aiyer2005bar} allows 
for both Byzantine, Acquiescent (honest) and Rational processors in the same 
system. They assume strong cryptographic primitives (bounded 
computation limits), and local approximately synchronized clocks for the 
processors. They build a distributed backup system for rational agents, that is 
resilient up to $t<\frac{1}{3}n$ Byzantine processors that wish to minimize the 
utility function of the rational agents, and is resilient to a deviation of a 
single rational agent (i.e., 2 rational agents might be able to collude and 
enlarge their utility).

In \cite{abraham2006distributed}, Abraham et al. introduce the term of 
resilience in the way we use it. They study secret sharing and multi-party 
computation in a synchronous fully connected network, while assuming rational 
players want to learn the secret but also want as few as possible of the other 
players to learn the secret. They provide an solution, based on Shamir's secret 
sharing scheme that is resilient to coalitions of size $n-1$. Further, they 
they apply their methodology to simulate mediators, and to support the 
deviation of malicious processors.

As a complementary work, in \cite{abraham2008lower}, Abraham et al. present 
lower bounds for implementing a mediator using message passing (cheap talk), in 
a synchronous fully connected network.

There is a variety of game-theoretic approaches to distributed
computing. A discussion about the basic definitions and a brief
survey can be found in \cite{abraham2011distributed}. A well studied
problem in the intersection of game-theoretic and distributed
computing is secret sharing and multi-party computation
\cite{halpern2004rational,abraham2006distributed,kol2008games,
	dani2011scalable,fuchsbauer2010efficient,gordon2006rational,lysyanskaya2006rationality}.
Recall that our main procedure is a sub-protocol that performs
secret sharing.

One of the early studied models for resilient Fair leader election, or fair 
coin toss (which are usually equivalent) was the full information model 
suggested by Ben-Or and Linial \cite{}. Assuming each processor plays in its 
turn, by broadcasting a message to all the processors, fair coin toss was 
studied in
\cite{saks1989robust,ajtai1993influence,
	alon1993coin,zuckerman1996randomness,boppana2000perfect,russell2001perfect}.
A protocol is an extensive game with perfect information. Each player 
(processor) has an unlimited computation power. Each player in its turn 
broadcasts its current action. Saks \cite{saks1989robust} suggested pass the 
baton, a fair leader election protocol that is resilient to coalitions of size 
$O(n / \log(n))$ . In \cite{ben1990collective,ajtai1993influence}, Ben-Or and 
Linial and Atjai et al. studied a certain class of full information coin toss 
games, which can be expressed by  $n$ variable boolean functions. They showed 
that in their games, $n / \log^2(n)$ players can bias the output.

In \cite{alon1993coin}, Alon et al. showed that a random protocol achieves is 
resilient to coalitions of a linear size. Later, Boppana and Narayanan 
\cite{boppana2000perfect} proved the existence of such a protocol with near 
optimal resilience, that is, resilience to coalitions of size $(\frac{1}{2} - 
\epsilon)n$. Finally, \cite{russell2001perfect} presented a constructive 
protocol that gives $(\frac{1}{2} - \epsilon)$ resilience in time $\log^*(n)$ 
($n \log^*(n)$ messages - executed in $n$ asynchronous rounds).

Inspired by \cite{alon1993coin}, we construct a protocol that is based on a 
non-constructive random function.

Recently, Afek et al. \cite{afek2017cheating}, studied resilient protocols from 
another angle. They ask how much information about $n$ processors must have 
in order to implement a resilient protocol. They study this question in the 
message passing model, in synchronous networks of general topology.

\mySection{Model}
\label{section:Model}

We use an asynchronous version of the LOCAL computation model (see,
\cite{PelegBook}). That is, the processors are nodes on a
communication graph $G=(V,E)$ and they communicate by sending
messages of unlimited size along the edges. Messages are guaranteed to arrive
uncorrupted in a FIFO order.
Processors are allowed to perform computations and send messages only upon wake
up, or upon
receiving a message. Additionally, each processor may perform local
randomization. Equivalently, each processor has an infinite
\textit{random string} as input and it operates deterministically.
Each processor has a unique $id$ which it cannot modify. The set of
$id$s, $V$, is known to the processors, therefore w.l.o.g we may
assume that $V=[n]:=\{1,\ldots,n\}$. When a processor receives a
message, it may send zero or more messages and afterwards it may
also select some $output$ and terminate. The $output$ may be any
value, including $\bot$ which denotes \textit{abort}. The messages
are delivered asynchronously along the links by some oblivious
message schedule which does not depend on the messages' values.

A \emph{\textbf{strategy}} of a processor is a (deterministic)
function that defines its behavior. Upon waking-up or receiving an
incoming message, the strategy decides what messages to send and
whether or not to terminate. The decision is based on everything
known to the processor until that time: Its $id$, its random string
and its history (all the messages it has received).  A
\textbf{protocol} is a vector of $n$ strategies - a strategy for
each processor in $ V $. A \emph{\textbf{symmetric protocol}}, is a
protocol that provides the same strategy to all the processors. In
game-theoretic terms, the processors are the players and a protocol
is a strategy profile.

Given an execution $e$ of a protocol, define $outcome(e)=o$ (for
some $o\in V$) if all processors terminate with $output=o$. We call
such an outcome $o\in V$, {\em valid}. Otherwise, if either some
processor never terminates, or some processor $i$ terminates with
$output_i=\bot$, or some processors $i$ and $j$ terminate with
$output_i\neq output_j$, then we have $outcome(e)=\FAIL$. Notice
that the $output$ of each processor is determined \textit{locally},
while the $outcome$ of an execution is a function of all the
individual outputs so is therefore determined \textit{globally}.

The solution preference assumption might seem problematic due to
this definition of outcome. At first glance, one might think that a
cheater could ``force'' all processors to agree on its preferred
outcome by always terminating with its most preferred output. If all
players know this behavior, since they prefer any valid outcome over
a failure, then they will aline with the cheater. However, it is not
the case because the strategy of each honest (non-cheating) agent is
predetermined. That is, the agents do not have any side-channel to
discuss threats. As a motivating reasoning, the technician installs
the program on each computer and it is never modified.

A \textbf{fair leader election} (\textbf{FLE}) protocol $P$ elects a
leader uniformly. Formally, $P$ is a symmetric protocol that assigns
a strategy $S$ to every processor such that for every message
schedule
    \[\forall j\in V: Pr(outcome(e)=j)=\frac{1}{n},\]
where the probability is over the local randomization of the
processors.

In order to define a game, we assume that each processor maximizes
their expected utility, which is only a function of the outcome.
More, we assume that each processor is rational,
i.e., it has a higher utility for valid outcomes. Formally,
\begin{definition}
A \emph{\textbf{rational utility}} of a processor $ p $ is a
function $u_p: [n] \cup \{\FAIL\} \to [0, 1]$, such that $u_p(\FAIL)
=0$.
\end{definition}

The motivation for the definition is that each processor, including
the deviating processors, would prefer any legitimate outcome (in
$V$) over any other outcome (which will result in $\FAIL$), i.e., we assume the
solution preference assumption.
%
%
Notice that any processor can force $outcome(e)=\FAIL$ by aborting
(terminating with $output=\bot$). So if we had $u_p(\FAIL) > u_p(i)$
for some $i \in [n]$ then whenever $p$ sees that the output is going
to be $i$, it would simply abort instead. Intuitively, processors
would like to promote their preferred leader while having the
protocol succeed.

We start by defining
a deviation of a coalition.
\begin{definition}{(Adversarial Deviation)}
Let $P$ be a symmetric protocol that assigns the strategy $S$ to
every processor. Let $C\subset V$ be a subset of $k$ processors. An
\textit{adversarial deviation} of $C$ from $P$ is a protocol $P'$,
in which every processor $i\not\in C$ executes $S$ and every
processor $i\in C$ executes an arbitrary strategy $P'_i$. The
processors in $C$ are called \textit{adversaries} and the processors
not in $C$ (i.e., in $V\backslash C$) are called \textit{honest}.
\end{definition}
Concisely, a protocol is {\bf \kresilient} if no coalition of size
$k$ can increase the expected utility of each of its members by at
least $\epsilon$ by an adversarial deviation (note that this is an
$\epsilon$-$k$-Strong Nash equilibria). A protocol is
\perfectkresilient if it is \kresilient for $\epsilon=0$. Formally,
\begin{definition}
A protocol $P$ is {\bf \kresilient} if, for every oblivious messages
schedule, for every rational utilities, for every coalition $C$ of
size $k$, and for every adversarial deviation $D=(P_{V-C}, P'_{C})$
of the coalition $C$ using $P'$, there exists  $ p\in C$ such that,
\[E_D[u_p] \leq E_P[u_p] + \epsilon\]
\end{definition}
For a unidirectional ring, which is the focus of this paper, all
message schedules are equivalent because each processor has only one
incoming FIFO link. For a general scenario, the above definition
implies that the adversaries may choose any oblivious schedule. But
the selection of the schedule may not depend on the inputs or on the
processors' randomization.

To simplify the proofs, rather than considering the expected utility
of each adversary, we consider the change in probabilities of valid
outcomes. An FLE protocol $P$ with is {\it \kunbiased} if for every
adversarial deviation $D$ of size $k$:
        \[\forall j\in V: Pr_D(outcome(e)=j) \leq \frac{1}{n}+\epsilon\]

The following lemma shows the equivalence of resilience and unbias.
\begin{restatable}{lemma}{RestatableUnbiasResilienceEquiv}
    \label{Model:UnbiasResilienceEquiv}
If an FLE protocol $P$ is \kresilient then it is \kunbiased. If an FLE
protocol is \kunbiased then it is $(n\epsilon)$-$k$-resilient.
\end{restatable}
\begin{proof}
    Let $P$ be an \kresilient FLE protocol ans $C$ be an adversarial coalition of size $k$.
    Assign the
    following rational utility to every processor $ p\in C$ we have $u_p(j): =
    \mathbbm{1}_{[j=j_0]}$ and for $p\not\in C$ we have $u_p(j):=\mathbbm{1}_{[j=p]}$.
    (We can select any utility 
    for $p\not\in C$ and the same proof holds.)
    Let $D$ be an adversarial deviation from $P$ for $C$.
    Then by resilience we get, for $p\in C$, $E_D[u_p] \leq E_P[u_p] + \epsilon =
    \frac{1}{n} + \epsilon$, but $E_D[u_p]=Pr_D(outcome=j_0)$, so
    $Pr_D(outcome=j_0)\leq \frac{1}{n}+\epsilon$. Therefore $P$ is \kunbiased.

    For the other direction, let $P$ be an \kunbiased FLE protocol. Fix a processor $p$ and let $u_p$ be its rational utility.
    Let
    $D$ be an adversarial deviation of size $k$. Since $P$ is unbiased, we get
    $\forall j\in [n]: Pr_D(outcome=j) \leq \frac{1}{n} + \epsilon$.
    So,
    $E_D[u_p] = \sum_{j\in[n]} Pr_D(outcome=j) u_p(j) \leq \sum_{j\in[n]}
    (\frac{1}{n} + \epsilon) u_p(j) = $\\
    $\sum_{j\in[n]} Pr_P(outcome=j) u_p(j) +
    \sum_{j\in[n]} \epsilon u_p(j)
    \leq E_P[u_p] + \epsilon n$. Therefore $P$ is
    $(n\epsilon)$-$k$-\textit{resilient}.
\end{proof}

\mySection{A Resilient Fair Leader Election Protocol for an Asynchronous 
Unidirectional Ring}
\label{section:OriginalProtocol}

We present \RingOriginalProtocol, the asynchronous unidirectional
ring FLE protocol of
\cite{abraham2013distributed,afek2014distributed}. The protocol
relies on a \textit{secret sharing} sub-protocol. First, we describe
the protocol without specifying the implementation of the secret
sharing sub-protocol. Then we present its implementation.

Each processor $i$, selects a secret $d_i\in [n]$ uniformly. Then,
using a secret sharing sub-protocol, all processors share the secret
values $\DataValues$ with each other, such that each processor $i$
gets the values $\hat{d}_{i,1}, \hat{d}_{i,2}, ...\hat{d}_{i,n} $
where $\hat{d}_{i,j}=d_j$ for all $j$. Then, each processor $i$
validates locally that $\hat{d}_{i,i}=d_i$. If $\hat{d}_{i,i} \ne
d_i$ then it aborts by terminating with $output_i=\bot$. Finally,
each processor $i$ terminates with $output_i=\sum_{j=1}^n
\hat{d}_{i,j} \pmod n$.

It remains to define the secret sharing sub-protocol. For didactic
reasons, first consider the following non-resilient secret-sharing
sub-protocol as in \cite{abraham2013distributed}: Each processor $j$
sends its secret $d_j$, and then forwards $n-1$ messages (receives
and sends immediately). If all processors execute this sub-protocol
honestly, then each processor receives every secret exactly once.
Using the scheme defined above with this secret sharing sub-protocol
is not resilient even to a single adversary (a coalition of size
$k=1$). An adversary could wait to receive $n-1$ values before
sending its first message and then select its secret value to
control the total sum $\SumDataValues$. (The pseudo-code can be
found in \Variant{the full
paper}{Appendix~\ref{section:BasicNonResilientProtocol}}.)

Ideally, we want every processor to ``commit'' to its secret value
before knowing any other secret value. In order to force processors
to ``commit'' to their values, the processors delay every incoming
message for one round. W.l.o.g., define processor $1$ to be the
\textit{origin} processor, and define it to be the only processor
which wakes up spontaneously. Let the rest of the processors be
\textit{normal} processors. We specify different functionality for
the \textit{origin} processor and for the \textit{normal}
processors.

\medskip

\begin{algorithm}[h]
    \caption{Secret sharing for \RingOriginalProtocol, shares the values $\DataValues$}
\textit{Strategy for a normal processor $i$}: Initially, store $d_i$
in a buffer. For the following $n$ incoming messages, upon receiving
a new message $m$, send the value which is currently in the buffer
and then store $m$ in the buffer.

\textit{Strategy for origin}: Initially (upon wake-up) send $d_1$
and then forward (receive and send immediately) $n-1$ incoming
messages.
\end{algorithm}\medskip

The artificial delay in the secret sharing defined above forces
every processor to give away its secret value before it gets to know
the other secret values. Furthermore, this delay limits the
communication of the adversaries. Two adversaries that are separated
by $l$ consecutive honest processors need to send $l+1$ messages in
order to transfer information.

\begin{figure}
    \centering
    \includegraphics[width=0.35\textwidth,keepaspectratio]{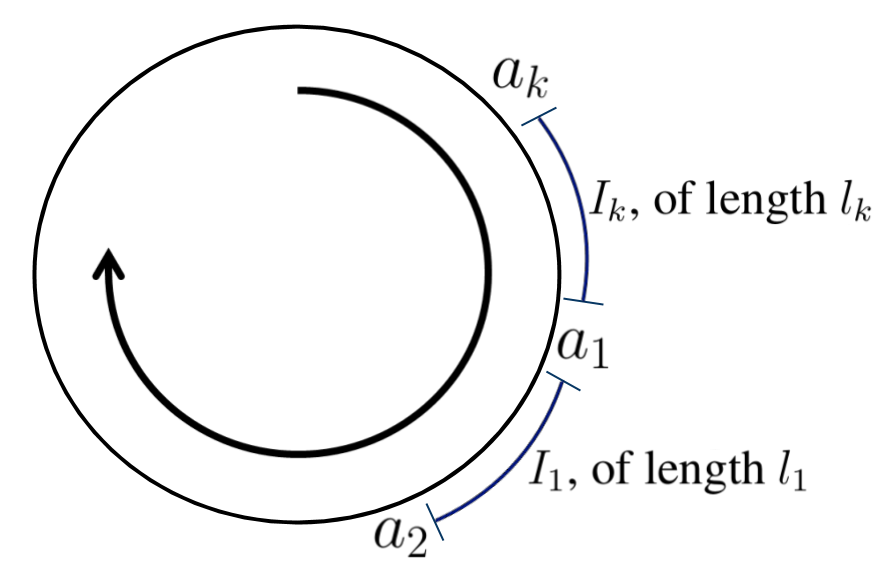}
    \caption{The adversaries locations on the ring.}
   \label{Figure:Ring}
\end{figure}

\begin{definition}[\textit{honest segment}]
Given an adversarial coalition $C=\{a_1, \ldots, a_k\}\subseteq V$,
where there is no adversary $a_i$ between $a_j$ and $a_{j+1}$, a
maximal set of consecutive honest processors is called an
\textit{honest segment}. Denote by $I_j \subseteq V$ the honest
segment between $a_j$ and $a_{j+1}$, and let $l_j$ be its length.
(See Figure \ref{Figure:Ring}.)
\end{definition}
\begin{definition}
An adversary $a_i$ with non-trivial
segment $l_i \geq 1$ is called an \textit{exposed adversary}.
\end{definition}

By the definition of $outcome$, an execution of an adversarial
deviation from \RingOriginalProtocolSpaced might have $outcome=\FAIL$ for
three reasons. First, the execution might run forever because some
exposed adversary sends less than $n$ messages. Second, two honest
processors might calculate different outputs. Third, an honest
processor $h$ might abort by terminating with $output=\bot$ if its
$n^{th}$ incoming message is invalid, i.e., $\hat{d}_{h,h}\ne d_h$.
We characterize these reasons in the following lemma.

\begin{restatable}{lemma}{BasicValidExecution}
    \label{Basic:ValidExecution}
For every adversarial deviation from \RingOriginalProtocol, an
execution succeeds ($outcome \ne \FAIL$) if and only if the
following conditions hold:
\begin{enumerate}[noitemsep,topsep=0pt]
\item
Every exposed adversary sends $n$ messages.
\item
The sums of all the outgoing messages of the exposed adversaries,
are identical modulo $ n$.
\item
For every adversary $a_j$, its last $ l_j $ messages are the secret
values of the honest processors in $ I_j $, in the appropriate
order.
\end{enumerate}
\end{restatable}
Note that in order to bias the output, the adversaries do not
necessarily need to send the same set of messages. They only need to
comply to conditions $1$ and $3$ while controlling the sum of their
outgoing messages (in addition to condition $2$). In our attacks we
show this is indeed possible.
Note that condition $1$ implies that every honest processor sends
$n$ messages and therefore all the honest processors terminate.

\begin{lemma}\label{SumOutgoingLemma}
    For every $ a_j \in C$, if the last $ l_j $ outgoing messages of $ a_j $
    are the secret values of $ I_j $ in the appropriate order, then the
    calculated sum of every processor in $ I_j $ is the sum of the outgoing
    messages of $ a_j $.
\end{lemma}
\begin{proof}
    Assume that the last $ l_j $ messages of $ a_j $ are the random values of $
    I_j $ in the appropriate order. Since $I_j$ is continuous, it is enough to
    show that every two consecutive processors in $I_j$ calculate the same
    sum.\\
    Let $ b $, $ c $ be two consecutive processors in $ I_j $ where $ c $ is
    the successor of $ b $.
    Let $ [m_1, m_2, ..., m_n] $ be the incoming messages of $ b $. Since $ b $
    is honest, the incoming messages of $ c $ are $ [r, m_1, ...,$ $ m_{n-1}] $
    where $ r $ is the random value of $ b $. Since the last $ l_j $ messages
    of $a_j$ are the random values of $ I_j $, the last message that $ b $
    receives, $ m_n $, is its random value $ r $. Therefore, $b$ and $c$
    received the same list of messages up to a permutation, in particular their
    calculated sums in $e$ are equal.
\end{proof}

\begin{lemma}\label{ValidationPassLemma}
    Let $e$ be an execution of $P$. Let $ a_j \in C$. The last $ l_j $ messages
    of $ a_j $ are the random values of $ I_j $ in the appropriate order, if
    and only if all the processors in $I_j$ pass validation\Variant{}{on line
    \ref{alg:validateLastValue2}}.
\end{lemma}
\begin{proof}
    Denote $I_j=(h_1, h_2, ..., h_{l_j})$ the honest processors along $I_j$,
    denote $(m_{l_j}, ..., m_2, m_1)$ the last $l_j$ messages of $a_j$ in the
    order they are sent. The last message that $h_1$ receives is $a_j$'s last
    outgoing message, $m_1$. Since $h_1$ is honest, the last message that $h_2$
    receives is $m_2$ etc. So every processor in $I_j$ receives its random
    value as the last message if and only if the last $ l_j $ messages of $ a_j
    $ are the random values of $ I_j $ in the appropriate order.\\
    We conclude because a processor passes validation\Variant{}{at line
    \ref{alg:validateLastValue2}} if and only if its last incoming message is
    its random value.
\end{proof}

\BasicValidExecution*
\begin{proof}
    ($\Rightarrow$)Let $e$ be an execution of $P$. Assume that either 1 or 2
    does not hold. If 1 does not hold, then let $ a_i$, $ a_j $ be two such adversaries with
    different sum of outgoing messages $ S_i, S_j $. Let $ b_i, b_j $ be their
    successors, they are honest because $l_i>0$ and $l_j>0$. So $b_i$ elects
    $L[S_i]$ and $b_j$ elects $L[S_j]$. There are two different outputs, and
    therefore $outcome=\FAIL$.\\
    If 2 does not hold, then one of the processors in $I_j$ fails validation by
    Lemma \ref{ValidationPassLemma}, it aborts and therefore $outcome=\FAIL$.

    ($\Leftarrow$)Let $e$ be an execution of $P$. Assume that 1 and 2 hold.
    Since 2 holds, all processors pass validation\Variant{}{at line
    \ref{alg:validateLastValue2}}, therefore they all output a valid value in
    $[n]$ ($output\ne\bot$). So it is enough to show that all the honest processors calculate the same
    sum. Let $ h_i\in I_i $, $ h_j\in I_j $ be two honest processors. Since 2
    holds, from Lemma \ref{SumOutgoingLemma} the sum that $ h_i $ calculates is
    the sum of outgoing messages of $ a_i $, similarly for $ h_j $. Since 1
    holds, these two sums are equal, so $ h_i $ and $ h_j $ calculate the same
    sum. Therefore all the honest processor terminate with the same output.
\end{proof}

\textbf{Remark}. Originally, in the model defined in
\cite{abraham2013distributed}, the $ids$ are unknown prior to the
execution, so the protocol begins with a wake-up phase, in which
processors exchange $id$s and select an orientation for the ring.
Clearly, our attacks still hold for the original protocol, since the
adversarial processors can behave honestly during this initial
phase.
We are unsure how to extend our resilience proofs to handle this
case. The worry is that adversaries can abuse the wake-up phase in
order to transfer information.

\textbf{Remark}. There exist general commitment schemes in
other research areas, but they do not fit in our model. Since we assume
unlimited computation power, generic
computation-based cryptographic commitment schemes such as one-way functions
are useless in our model.

\mySection{Adversarial Attacks on \RingOriginalProtocolTitle}
\label{section:Attacks}

In this \sectionText we describe the adversarial attacks on
\RingOriginalProtocol. First, we show that a coalition of size
$k=\sqrt{n}$ located at equal distances \textbf{can control the
outcome}. Namely, for any $w\in [n]$ they can force $outcome=w$.
Second, we show that, with high probability, a coalition of
$O(\sqrt{n \log n})$ randomly located processors can control the
outcome. Third, we show that an adversarially located coalition of
size $O(\sqrt[3]{n})$ can also control the outcome.

The case of equally spaced coalition of size $k=\sqrt{n}$ follows
from the following.
\begin{lemma}
    \label{Attacks:NaiveAttackLemma}
For every coalition $C\subseteq V$ such that every honest segment
$I_j$ is of length $l_j \leq k-1$, the adversaries can control the
outcome. I.e., for every $w \in [n]$, there exists an adversarial
deviation $D$ from \RingOriginalProtocolSpaced such that for every
execution $e$ of $D$: $outcome(e)=w$.
\end{lemma}
\begin{proof}
We show that the adversaries can comply to conditions 1 and 3 of
Lemma \ref{Basic:ValidExecution} while controlling the sum of
outgoing messages of every adversary. The main idea is that
adversaries never select a secret value for themselves. Moreover,
instead of buffering every incoming message, the adversaries just
\textit{forward every incoming message immediately}. This way, after
$n-k$ rounds, every adversary sent only $n-k$ messages and knows all
the secret values of all the $n-k$ honest processors.

Each adversary $a_j$ can control the sum of its outgoing messages
while complying to conditions $1$ and $3$ of
Lemma~\ref{Basic:ValidExecution}: It sends a message $M$ (we explain
later how to choose $M$), then it sends $k-l_j-1$ times $0$, and
finally it sends its last $l_j$ secret value messages of the honest
processors in $I_j$, as expected. Since the total sum is $\Gamma =
\sum_{i\notin C} d_i + M + 0\cdot(k-l_j-1)+ \sum_{i\in I_j}d_i \pmod
n$, adversary $a_j$ can control this sum by selecting $M$ properly,
i.e., for $M=w - \sum_{i\notin C} d_i - \sum_{i\in I_j}d_i\pmod n$
we have $\Gamma=w$.
\end{proof}

From the above lemma we deduce the following theorem.
\begin{theorem}
    \RingOriginalProtocolSpaced is \textbf{not} \kresilient for every $k \geq
    \sqrt{n}$, $\epsilon < 1 - \frac{1}{n}$.
\end{theorem}
\begin{proof}
Let $C$ be a coalition of size $k \geq \sqrt{n}$ located at equal
distances along the ring (equal distances means $|l_i-l_j| \leq 1$).
Every honest segment $I_j$, is of length $l_j <
\frac{n-k}{k}+1=\frac{n}{k} \leq \sqrt[]{n} \leq k$. So, $l_j \leq
k-1$. Therefore the condition for Lemma
\ref{Attacks:NaiveAttackLemma} holds and the adversaries can control
the outcome, i.e., $Pr(outcome=1) = 1 = \frac{1}{n} + (1 -
\frac{1}{n})$. Therefore \RingOriginalProtocol is not \kunbiased for
$\epsilon < 1 - \frac{1}{n}$, and, by Lemma
\ref{Model:UnbiasResilienceEquiv}, it is not \kresilient for
 $\epsilon < 1 - \frac{1}{n}$.
\end{proof}

Notice that Lemma \ref{Attacks:NaiveAttackLemma} requires only $l_j \leq
k-1$ for all $j$. Unsurprisingly,
$k=\Theta(\sqrt{n \log n})$ \textit{randomly} located
adversaries, with high probability, will comply to this
requirement (an explicit calculation is included in \Variant{the full
paper}{Appendix \ref{section:RandomizedModel}}). Therefore,
\RingOriginalProtocolSpaced is not resilient
against $k=\Theta(\sqrt{n \log n})$ randomly located adversaries. In
\Variant{the full paper}{Appendix \ref{section:RandomizedModel}} we also show a
similar attack for
$k=\Theta(\sqrt{n \log n})$ randomly located adversaries that do not
even know their distances $\{l_j\}$ and the exact number of
adversaries $k$.

Next, we improve the attack from Lemma
\ref{Attacks:NaiveAttackLemma} and show that $k=\Theta(\sqrt[3]{n})$
adversaries can control the outcome. The key observation from Lemma
\ref{Attacks:NaiveAttackLemma} is that the adversaries do not need
to select a secret value for themselves so they can transfer the
secret values of the honest processors faster than expected.
Notice that when the adversaries do not send their values,
they have $k$ extra messages they are allowed to send. In the new attack, the
adversaries leverage these extra messages to ``push'' information
faster along the ring.

Technically,  we locate the $k$ adversaries having the following
distances $l_i = (k+1-i)(k-1)$. For simplicity, one can think that
$n=k+(k-1)\sum_{i=1}^{k}i=\frac{1}{2}k^3+\Theta(k^2)$. However, we prove the 
attack works for general $k$ and $n$ such that $k \geq
2\sqrt[3]{n}$. We show that a coalition, with such distances, can control the
output.\medskip

\begin{algorithm}[h]
    \caption{Cubic Attack, strategy for adversary $a_i$, for electing $w$.}
    \begin{enumerate}[noitemsep,topsep=0pt]
        \item \label{CubicPhase:Transfer} Transfer (receive and send immediately) $ n - k - l_i $ incoming messages.\\
                Denote with $ m_j $ the $ j^{th}$ message that was received.
        \item \label{CubicPhase:SendZeros} Send $ k-1 $ messages with the value $0$.
            \item \label{CubicPhase:Wait} Wait to receive $l_i$ more incoming messages, to get a total of $n-k$ messages\\
                (only receive, do not send them).
        \item \label{CubicPhase:CancelSum} Send the message $M=w-\sum_{j=1}^{n-k} m_j \pmod n$.
        \item \label{CubicPhase:SendLast} Send the following messages, one after the other, $m_{n-k-l_i+1}, \ldots , m_{n-k}$
    \end{enumerate}
\end{algorithm}
See an explicit pseudo-code in \Variant{the full paper}{Appendix 
\ref{section:CubicAttackPseudoCode}}.
\medskip

\begin{restatable}[Cubic Attack]{theorem}{CubicAttackTheorem}
    \label{Attacks:CubicAttackTheorem}
    \RingOriginalProtocolSpaced is \textbf{not} \kunbiased for every $\epsilon <    1-\frac{1}{n}$ and $k \geq 2\sqrt[3]{n}$.
\end{restatable}

For simplicity, assume that the origin is honest. Recall
we assumed the distances are $\forall i: l_i=(k+1-i)(k-1)$. In this section,
we relax this requirement to  $l_k \leq k-1, \forall i<k: l_i \leq l_{i+1} +
k-1$.

Assume we have $k' > 2\sqrt[3]{n}$ adversaries, calculation shows
that for every $n>4$: $\sum_{i=1}^{k'-1}
(k'-1)i=(k'-1)\frac{k'(k'+1)}{2} \geq n - k'$. Let $k \leq k'$ be
the minimal integer such that $\sum_{i=1}^{k-1} (k-1)i \geq n - k$.
Let $(l_i)_{i=1}^k$ be integers such that $\forall i: l_i \leq
l_{i+1} + k-1$ and $l_k \leq k-1$. Locate $k$ adversaries along the
ring within distances $(l_i)_{i=1}^k$ where $l_i$ denotes the
distance between $a_i$ and $a_{i+1}$. Locate the rest of the $k'-k$
adversaries arbitrarily and define them to behave honestly.

\begin{lemma}
    \label{CubicAttack:AllTerminate}
    All adversaries terminate.
\end{lemma}
\begin{proof}
    Since $l_1 = max_i(l_i)$, for $ n - k - l_1$ rounds all the adversaries
    behave like pipes. So after $n-k-l_1$ rounds, for every $i$, $a_i$ received
    and sent $n-k-l_1$ messages.\\
    Then, $a_1$ begins step \ref{CubicPhase:SendZeros}. It sends $k-1$ zeros.
    So $a_2$ received $n-k-l_1+k-1 \geq n-k-l_2$ messages. Then $a_2$ begins
    step \ref{CubicPhase:SendZeros}. Now $a_2$ sends $k-1$ zeros, therefore
    $a_3$ begins step \ref{CubicPhase:SendZeros} and so on. Until $a_k$
    completes step \ref{CubicPhase:SendZeros}.\\
    After completing step \ref{CubicPhase:SendZeros}, $a_k$ sent $n-k-l_k+k-1
    \geq n-k$ messages. Therefore, $a_1$ receives a total of at least $n-k$
    messages, so it completes waiting in step \ref{CubicPhase:Wait}, performs
    steps \ref{CubicPhase:CancelSum}-\ref{CubicPhase:SendLast} and terminates.
    Then,  $a_2$ receives $n \geq n-k$ messages, so it completes as well and
    then $a_3$ completes and so on.
\end{proof}
Let $I_i = \{a_i+1, ..., a_i+l_i\}$ be an honest segment. Denote the
reversed series of the secret values of the segment $secret(I_i) =
(d_{a_i+l_i}, ... d_{a_i+1})$.
\begin{lemma}
    \label{CubicAttack:SecretValueOnly}
    In the Cubic Attack, the first $n-k$ incoming messages of each adversary processor $a_i$ are the
    secret values of the honest processors according to their order along the
    ring, that is
    $secret(I_{i-1}),$ $secret(I_{i-2}),$ $secret(I_{i-3}),$ $..., secret(I_{i})$
\end{lemma}
\begin{proof}
    For each adversary $a_i$, the first $l_{i-1}$ incoming messages of $a_i$
    are $secret(I_{i-1})$. Its following $l_{i-2}$ incoming messages are the
    first $l_{i-2}$ outgoing messages of $a_{i-1}$ which are the first $l_{i-2}$ incoming
    messages of $a_{i-1}$, which are $secret(I_{i-2})$ and so on. So the first $n-k =
    \sum_{j=1}^{k} l_{i-j\pmod k}$ incoming messages of $a_i$ are
    $secret(I_{i-1}) secret(I_{i-2}), secret(I_{i-3}),..., secret(I_i)$.
\end{proof}
Notice that we implicitly used the fact that each adversary $a_i$
transfers its first $n-k-l_i$ incoming messages. A careful proof by
explicit induction uses it.

\CubicAttackTheorem*
\begin{proof}
    Consider the adversarial deviation described above. By Lemma
    \ref{CubicAttack:AllTerminate}, it is enough to show that all the honest
    processors terminate with $output=t$. Therefore, by Lemma
    \ref{Basic:ValidExecution}, it is enough to show that for each adversary
    $a_i$, the last $l_i$ outgoing messages of $a_i$ are the secret values of
    its honest segment $I_i$, and that the total sum of its outgoing messages
    is $t$.\\
    By Lemma \ref{CubicAttack:SecretValueOnly}, the first $n-k$ incoming
    messages of $a_i$ are $secret(I_{i-1}), secret(I_{i-2}), secret(I_{i-3}),
    ..., secret(I_{i})$. In particular, the last $l_i$ of them $m_{n-k-l_i+1},
    ..., m_{n-k}$ are $secret(I_i)$. So the last $l_i$ outgoing messages of
    $a_i$ are the secret value of $I_i$ as required.\\
    More, let us calculate the sum of outgoing messages of $a_i$. By
    definition, the sum is $\sum_{j=1}^{n-k-l_i} m_j + (t-S) +
    \sum_{j=n-k-l_i+1}^{n-k} m_j = t - S + S = t$ as required.
\end{proof}

\begin{corollary}
    \RingOriginalProtocolSpaced is \textbf{not} \kresilient for every $0 \leq
    \epsilon \leq 1-\frac{1}{n}, k \geq 2\sqrt[3]{n}$
\end{corollary}

Finally, we conjecture that \RingOriginalProtocolSpaced is
\perfectkresilient for $k=\Omega(\sqrt[3]{n})$:
\begin{conjecture}
    There exists a constant $\alpha>\frac{1}{8}$, such that for every large enough $n$,  \RingOriginalProtocolSpaced is \perfectkresilient for every $k \leq \alpha\sqrt[3]{n}$.
\end{conjecture}

\mySection{Resilience Results for \RingOriginalProtocolTitle}
\label{section:GoodBounds}

In this section, we outline a proof that shows
\RingOriginalProtocolSpaced is \kresilient for
$\epsilon=n^{-\Omega(\sqrt[4]{n})}$ and $k=O(\sqrt[4]{n})$. (See a
complete proof in \Variant{the full paper}{Appendix 
\ref{section:ALeadIsResilient}}.)
%
%
For simplicity, assume the $origin$ is an adversary, which changes
the resilience bound by only $1$.
%
Our main resilience result
for \RingOriginalProtocolSpaced is:
\begin{restatable}{theorem}{RestatableMainUnbiasTheorem}
    \label{Quad:MainUnbiasTheorem}
For every $n$, \RingOriginalProtocolSpaced is \kresilient for all $ k\leq
k_0 := \frac{1}{4}\sqrt[4]{n}$ and $\epsilon\geq n^3n^{-k_0}$.
\end{restatable}
%

The proof of Theorem \ref{Quad:MainUnbiasTheorem} is in \Variant{the full
paper}{Appendix
\ref{section:ALeadIsResilient}}. Here we provide the intuition for three
observations that are the main ingredients in the proof.
Let $C=(a_1, ..., a_k)$, where $a_i\in[n]$ be an adversarial
coalition of size $k$.
For every adversarial coalition there exists an honest segment of
length at least $\frac{n-k}{k} > 60k^3$. W.l.o.g. assume that $a_1$
precedes that honest segment of length $l_1 \geq 60k^3$.

The first observation is that at every time point, the total number
of messages sent by two adversaries $a_i$ and $a_j$ is similar.
We show that from the outgoing messages of every adversary, it
should be possible to recover $n-k$ secret values of the honest
processors in order to pass all the validations. Since an adversary
sends only a total of $n$ messages, then it can send up to $ k $
spare messages - otherwise some adversary $ a_i $ will not be able
to send the last $ l_i $ messages correctly. From this, we deduce \Variant{}{in
Lemma \ref{Quad:SendLikeRecv}} that at every time point $t$, an
adversary cannot send $2k$ messages more than it has received by
time $t$. This implies \Variant{}{(Lemma \ref{Quad:SquareDiff})} that all
adversaries are approximately synchronized, i.e., the difference
between the total number of messages sent by $a_i$ and $a_j$ at any
time is at most $2k^2$.

The second observation is that $a_1$ needs to send at least $n-l_1$
messages before it can obtain any information about $d_{h_1}$, where
$h_1$ is the honest successor of $a_1$. In order for information
about $d_{h_1}$ reach $a_2$, $a_1$ must send at least $l_1$
messages. Then, in order for that information to travel from $a_i$
to $a_{i+1}$ for every $i$, the adversary $a_1$ must send at least
$l_i-4k^2$ more messages. Overall, $a_1$ sends at least $n-4k^3$
messages before any information about $d_{h_1}$ reaches $a_1$. From
the selection of $a_1$ we have $l_1>60k^3$, so
$n-4k^3>n-60k^3>n-l_i$ and the observation is complete.

The third observation, which is a direct result of Lemma
\ref{Basic:ValidExecution}, is that when the adversary $a_1$ sends
its $(n-l_1)^{th}$ outgoing message it commits to all of its
outgoing messages, since its last $l_1$ outgoing messages are
predetermined to be the secret values of $I_1$.

Combining the last two observations, we see that the only outgoing
message of $a_1$ which depends on $d_{h_1}$ is its last message,
which must be $d_{h_1}$, therefore the sum of its outgoing messages
distributes uniformly. So the output calculated by $h_1$ is
distributes uniformly.

\mySection{\PhaseAsyncLeadTitle - A New 
\texorpdfstring{$\epsilon$-$\Theta(\sqrt{n})$-resilient}{epsilon-sqrt{n}-resilient}
 Fair Leader Election Protocol}
\label{section:BetterAlgo}

In this section, we present \PhaseAsyncLead, a new FLE protocol,
which is based on \RingOriginalProtocolSpaced and show that it is
resilient to $k=O(\sqrt{n})$ adversaries. Our new protocol improves
upon \RingOriginalProtocol, for which $k=O(\sqrt[3]{n})$ adversaries
can control its outcome (\SectionText \ref{section:Attacks}).

As discussed before, \RingOriginalProtocolSpaced simulates ``rounds''. In
every round, each processor in its turn receives a data value and
then sends the data value it received in the previous round. Let
$Sent_i^t$ be the number of messages a processor $i$ sent until time
$t$. Without adversaries, in \RingOriginalProtocolSpaced we have for every
time  $t$, $|Sent_i^t-Sent_j^t| \leq 1$. This means that all
processors are ``\textit{synchronized}''.
The cubic attack utilizes the asynchronous nature of the network to
take the honest processors out of synchronization. Specifically, in
the cubic attack there exist an adversary $a_i$ and a time $t$, such
that $|Sent_i^t-Sent_1^t| = \Omega({k^2})$ and the honest processors
do not notice any deviation. The key observation that makes the
attack possible is that this gap $|Sent_i^t-Sent_1^t|$ is larger
than the longest honest segment $I_1$, therefore the adversary $a_1$
learns all the data values before committing, i.e., before sending
$n-l_1$ messages.

We modify the protocol \RingOriginalProtocolSpaced by adding a ``phase
validation'' mechanism that keeps all processors better synchronized.
That is, we enforce the following property: for every processor $i$
and time  $t$, $|Sent_i^t-Sent_1^t| = O(k)$.

The ``phase validation'' mechanism works as follows: Each processor
$i$ selects a secret \textbf{validation value} $v_i \in [m]$
uniformly (define $m=2n^2$). In round $i$, processor $i$ is the
current round \emph{validator}, and send $v_i$. All the other
processors transfer the validation value $v_i$ along the ring
without delay. Then, the round's validator validates that the
validation value it receives is indeed the same one it selected.

The random secret values in \RingOriginalProtocolSpaced are denoted by
$\DataValues$, and we call them \textbf{data values}. The $output$
in \RingOriginalProtocolSpaced is defined to be $\SumDataValues$. Recall
the point of commitment of adversary $a_j$: when an adversary $a_j$
sends its $n-l_j$ outgoing message. After the point of commitment,
the adversary $a_j$ is obligated to send the data values of the
honest segment $I_j$ and therefore cannot affect its outgoing
messages anymore. In \RingOriginalProtocol, as seen in Theorem
\ref{Quad:MainUnbiasTheorem},  there exists an adversary that cannot
find out the sum of the data values, $S:=\sum_{h\notin C}d_h \pmod
n$, before sending too many messages. Therefore, it commits to its
outgoing messages before being able to bias the output.

In \PhaseAsyncLead  every processor receives alternately a message
from the original protocol \RingOriginalProtocol, carrying a data
value, and a message  from the phase validation mechanism, carrying
a validation value. Therefore, each processor treats all the odd
incoming messages (first, third,  etc.) as \textbf{data messages}
and all the even incoming messages (second, fourth, etc.) as
\textbf{validation messages}.

While the phase validation mechanism described above keeps all the
processors synchronized, adding it to
\RingOriginalProtocolVerySpaced makes it non-resilient even to $k=4$
adversaries. The adversaries can abuse the validation messages to
share partial sums of $S=\sum_{h \notin C}d_h \pmod n$ quickly and
thus control the outcome. We give a full explanation of such an
attack in \Variant{the full paper}{\SubSomethingText
\ref{Sqrt:RandomFunctionIsRequired}}. We solve this problem by
substituting the ``sum'' function with a random function $f$, so
adversaries cannot calculate useful partial information about the
input.

The resilience proof of \PhaseAsyncLead relies on the disability of
adversaries to transfer enough information before committing. Due to
the difficulty in separating information about data values from
information about validation values, we apply $f$ not only on the
data values, but also on some of the validation values. We choose
the inputs to $f$ such that an adversary commits to them before
being able to bias $f$ by manipulating them. After sending $n-l_i$
data messages, $a_i$ is committed to all of its outgoing data
messages, however it could still manipulate its last $l_i$ outgoing
validation messages. Therefore, we apply $f$ only on the first $n-l$
validation messages where $l \leq \frac{n}{k} \leq
\max_j\{l_j\}=l_{j_0}$ (later we also lower bound $l$). This way,
$a_{j_0}$ is committed to all of its outgoing messages that affect
the output after sending only $n-l$ messages. Intuitively, after
$n-l$ rounds an adversary can collect $(n-l)$ honest validation
values and information about $(n-l+2k)$ data values, since it can
abuse $k$ validation values to collect information about data
values. We want this to be less than all the information that goes
into $f$. The total information that goes into $f$ is $(n-l) +
(n-k)$, so we want $(n-l) + (n-l+2k) < (n-l) + (n-k)$, therefore we
need $l
> 3k$. Combining this inequality with $l \leq \frac{n}{k}$ we deduce
that we need $k=O(\sqrt{n})$ and then we select $l =
\Theta(\sqrt{n})$. To conclude, we define $f \colon [n]^n\times
[m]^{n-l} \to [n]$ to be a fixed random function and the output
calculated by processor $i$ is $output_i=f(\hat{d}_{i,1},
\hat{d}_{i,2}, ..., \hat{d}_{i,n}, \hat{v}_{i,1}, \hat{v}_{i,2}, ...
\hat{v}_{i,n-l})$.

We define $l:=\lceil 10 \sqrt{n} \rceil $, because then for
$k<\frac{1}{10}\sqrt{n}$ there exists an honest segment of length at
least $l$. Assume w.l.o.g that $l_1 \geq l$, i.e., $a_1$ precedes a
long segment. So as soon as $a_1$ sends $n-l$ messages, it is
committed to the output.

Recall that \RingOriginalProtocolSpaced is composed from a scheme that
relies on secret sharing sub-protocol that shares $\DataValues$.\\
\PhaseAsyncLead is composed of a similar scheme which relies on a
stronger secret sharing sub-protocol: Each processor $i$, selects
secrets $d_i\in [n]$ and $ v_i\in[m]$ uniformly. Then, using a
secret sharing sub-protocol, all processors share the data values
and the validation values $\DataValues, \ValidationValues$ with each
other, such that each processor $i$ gets the values $\hat{d}_{i,1}$
and $ \hat{d}_{i,2}, ...\hat{d}_{i,n}, \hat{v}_{i,1}, \hat{v}_{i,2},
...\hat{v}_{i,n} $ where $\hat{v}_{i,j}=v_j$ and $\hat{d}_{i,j}=d_j$
for all $j$. Then, each processor $i$ validates locally its
identities, i.e., if $\hat{d}_{i,i} \ne d_i$ or  $\hat{v}_{i,i} \ne
v_i$ then it aborts by terminating with $output_i=\bot$. Finally,
each processor $i$ terminates with $output_i=f(\hat{d}_{i,1},
\hat{d}_{i,2}, ...,\hat{d}_{i,n}, \hat{v}_{i,1}, \hat{v}_{i,2}, ...
\hat{v}_{i,n-l})$.

As in \RingOriginalProtocol, processor $1$ is called $origin$ and
the rest of the processors are called $normal$ processors. For
notation simplicity, we assume the processors are located in an
ascending order along the ring $1,\ldots ,n$, however, our protocol
and resilience proof can be modified to cope with generally located
processors.\medskip

\begin{algorithm}[h]
    \caption{Secret sharing for \PhaseAsyncLead, shares the values
    $\DataValues$ and $\ValidationValues$}

    \textbf{Code for a $ normal $ processor $i$}:\\
    Initially, $\hat{d}_{i,i} := d_i$\\
    \For{$j = 1$ \KwTo $n$}{
        Wait to receive a data message $\hat{d}_{i, i-j \pmod n}$ and then send
        the previous data message $\hat{d}_{i, i-j+1 \pmod n}$.\\
        \uIf{$j \ne i$ }{
            wait for an incoming validation message $\hat{v}_{i,j}$ and forward
            it immediately.
        }\Else{
        perform validation by sending $v_i$ and waiting to receive it.
    }
}
\textbf{Code for $ origin $:}\\
Initially, $\hat{d}_{1,1} := d_1$\\
\For{$j = 1$ \KwTo $n$}{
    Send $\hat{d}_{1,n-j+2 \pmod n}$ and then wait to receive a data message
    $\hat{d}_{1,n-j+1 \pmod n}$ \\
    \uIf{$j \ne 1$ }{
        forward an incoming validation message $\hat{v}_{1,j}$\\
    }\Else{
    perform validation by sending $v_1$ and waiting to receive it.\\
}
}
send the last incoming data message $\hat{d}_{1,n-j \pmod n}$\\
\end{algorithm}

Next, is our main result.

\begin{restatable}{theorem}{SquareResilienceTheorem}
    \label{SquareResilience}
    With exponentially high probability over randomizing $f$, \PhaseAsyncLead
    is \kunbiased for every $\epsilon \geq n^{-\sqrt{n}}, k \leq
    \frac{1}{10}\sqrt{n}$
\end{restatable}

This result is asymptotically tight because with high probability
over selecting $f$, it is not \kresilient for $k=\sqrt{n} +
3$ because the rushing attack demonstrated in Lemma
\ref{Attacks:NaiveAttackLemma} works for \PhaseAsyncLead as well:
While rushing data messages and handling validation messages
honestly, within $n-k$ rounds, each adversary knows all the data
values, and the first $n-l<n-k$ validation values. Then, each
adversary can control at least 3 entries in the input of $f$. Thus,
for a random $f$, with high probability every adversary $a_j$ can
control the output of its segment $I_j$ almost for every input.

\subsomething{Proof outline}

As in \SectionText \ref{section:GoodBounds}, we perform the following
simplifications w.l.o.g: Consider only deterministic deviations and
notice the message schedule has no impact over the calculations. We
show that with high probability over $f$, $Pr(outcome=1) <
\frac{1}{n} + \epsilon$.

For deviations from \PhaseAsyncLead, the input space is $\chi :=
[n]^{n-k} \times [m]^{n-k}$ (recall that $m:=2n^2$). For a deviation $D$,
denote by
$NoFail^D$ the inputs for which every honest processor $h$ terminate
with a valid output, $ output_h \ne \bot $, i.e., inputs for every honest
processor $h$,
$\hat{d}_{h,h} = d_h$ and $\hat{v}_{h,h} = v_h$.

For every processor $b$, for every integer $1\leq i \leq 2n$, denote
by $send(b, i)$ the event that $b$ sends its $i^{th}$ outgoing
message. For every honest processor $h$, denote by $s(h)=send(h,2h)$
the event that $h$ sends its validation message as the validator.
Denote by $r(h)=send(h-1,2h)$ the event that its predecessor, $h-1$,
sends a message which is interpreted by $h$ as $\hat{v}_{h,h}$ (so
is expected to be equal to $v_h$). Define the event $nr :=
send(a_1,2(n-l))$ be the point of commitment of $a_1$. Notice that
after the event $nr$ occurs, $a_1$ is committed, i.e., it already
sent all the messages that affect the output as it is calculated by
processors in $I_1$, except for the data values of $I_1$ which are
predetermined anyway.

We write $\alpha \leadsto \beta$ if the event $\alpha$ happens before
the event $\beta$ for every message schedule (this is an intuitive definition,
in the full proof we give an equivalent, however more useful definition for
this notation).

Given a deviation and an input and an event $\alpha$, denote by
$data(\alpha)$ its value. An honest processor is
\textbf{unvalidated} if $data(s(h))$ is not used to calculate
$data(r(h))$ - for example, it is true if there exists a message
schedule such that $r(h)$ happens before $s(h)$ (in the full proof we
give an equivalent definition for this notation as well).

\begin{proofoutline}{Theorem \protect\ref{SquareResilience}}
    In this proof, we call an adversarial deviation a
    ``\emph{\textbf{deviation}}''.
    For each deviation, we partition the inputs space $\chi$ into three
    disjoint sets: $\chi=\chi_1^D \cupdot \chi_2^D \cupdot \chi_3^D$.

    Intuitively, the first set $\chi_1^D$, contains inputs for which the
    adversaries break synchronization severely before the event $nr$ occurs.
    Denote $M_0 = 2(n -l + 4k + 2)$. In an honest execution (no adversaries),
    we have the following linear order over the events $s(1) \leadsto s(2)
    \leadsto ... \leadsto s(n-l) \leadsto nr \leadsto s(n-l+1) \leadsto...
    \leadsto s(\frac{1}{2}M_0) \leadsto send(a_i, M_0) ... \leadsto s(n)$.
    Given a deviation, we say the synchronization is \textbf{broken severely}
    by an input, if there exists an adversary $a_i$ for which $send(a_i, M_0)
    \leadsto nr$. \Variant{Next, }{In Lemmas \ref{Sqrt:StoR} -
    \ref{Sqrt:OneOfThree}} we analyze
    the validation mechanism and show that when the linear order noted above
    does not hold, honest processors tend to be unvalidated. Leveraging these
    insights, \Variant{}{in the proof of Lemma \ref{Sqrt:AllUnvalidated},} we
    show that for
    inputs that break synchronization severely, there are at least $k+1$
    unvalidated honest processors. When an unvalidated processor is the round's
    validator, the adversaries ``guess'' the validation value because
    $data(r(h))$ is calculated independently of $data(s(h))$. \Variant{From
    that,}{In Lemma
    \ref{Sqrt:IgnoreChi1}} we deduce that $Pr(NoFail^D ~|~ \chi_1^D) \leq
    \frac{1}{n}$. Note that while the existence of a single unvalidated
    processor is enough for the explanation above, in the full proof we need $k
    + 1$ of them for deeper reasons.

    The second set, $\chi_2^D$, contains inputs for which the first $2(n-l)$
    outgoing messages of $a_1$ are not informative enough. For such inputs the
    adversaries are unlikely to reconstruct correctly the data values and the
    validation values in order for all the validations to succeed
    (namely, they need the following equalities to hold $\hat{d}_{i,i} = d_i$ and $\hat{v}_{i,i} = v_i$) and
    therefore the honest processors are likely to abort. More specifically, we
    get $Pr(NoFail^D ~|~ \chi_2^D) \leq \frac{1}{n}$\Variant{}{ in Lemma
    \ref{Sqrt:IgnoreChi2}}.

    The set $\chi_3^D$ includes the rest of the inputs.

    Notice that a deviation is defined by $2nk$ decision functions, that each
    receives a history (a list of incoming messages) and returns a list
    (possibly empty) of messages to send.

    We partition all the deterministic deviations into equivalence classes
    $[{}\cdot{}]_{\approx}$ according the first $M_0$ decision functions of
    each adversary. Intuitively, two deviations are equivalent if their
    behavior during the first $\frac{1}{2}M_0$ rounds is identical.

    \Variant{Next, }{In Lemma \ref{Sqrt:ChisAndGEquiv}} we see that the
    behavior of the deviation
    over inputs in $\chi_3^D$ until the point of commitment $nr$ is determined
    by the equivalence class of $D$ . From that we deduce that $Pr(outcome=1
    ~|~ \chi_3^D) > \frac{1}{n} + \epsilon$ implies a bias property over the
    class $[D]_{\approx}$\Variant{}{ in Lemma \ref{Sqrt:ReduceToEquivalence}}.
    Since $f$
    is random and since the first $2(n-l)$ outgoing messages of $a_1$ have many
    different options (by definition $\chi_3^D \cap \chi_2^D = \emptyset$, so
    they are ``informative''), \Variant{}{in Lemma \ref{Sqrt:BoundBinomial}} we
    deduce that
    the probability for that bias property to hold is low, by using a
    Hoeffding's concentration inequality. Then, applying a union bound over the
    equivalence classes we get that for a random $f$ it is likely that all
    classes do not have that bias property, which implies that $Pr(outcome=1
    ~|~ \chi_3^D) < \frac{1}{n} + \epsilon$ for every deviation $D$.

    From the law of total probability over $\{\chi_i^D\}_i$ we get that for
    every $D$: $Pr(outcome=1) \leq \frac{1}{n} + \epsilon$.
\end{proofoutline}

Full details and and proofs for \PhaseAsyncLead are available in \Variant{the
full paper}{Appendix \ref{section:PhaseAsyncLeadFull}}.

\mySection{Resilience Impossibility for Graphs which are Simulated by Trees}
\label{section:TreesImpossibility}

Abraham et al. \cite{abraham2013distributed} proved that for any
graph there is no \kresilient FLE protocol for $k\geq
\frac{1}{2}{n}$. In this \sectionText we generalize the result to graphs
that can be simulated by a tree, where each node in the tree
simulates at most $k$ processors. This is a generalization since
every graph can be simulated by a tree of size 2, where each
node in the tree simulates at most $\lceil \frac{1}{2}n \rceil$ processors.

\begin{definition}{($k$-simulated tree)}
An undirected graph $ G=(V,E) $ is a $k$-\textbf{simulated tree}, if
there exists a tree $T=(V_T,E_T)$ and a graph homomorphism $f: V \to V_T$
from $G$ to $T$ such that
\begin{enumerate}[noitemsep,topsep=0pt]
        \item For all $v\in V_T$, $|f^{-1}(v)| \leq k$
        \item For all $v\in V_T$, $f^{-1}(v)$ is connected in $G$.
\end{enumerate}
\end{definition}
Note that requiring $f$ to be a homomorphism means

$\{(f(x), f(y)) ~|~ (x,y)\in E\}
\subseteq E_T$. The mapping $f$ can be viewed as a partition of the
vertices  of $G$ to sets of size at most $k$,
such that each part is connected and the induced graph over the
partition constitutes a tree.

We show (in \Variant{the full paper}{Appendix 
\ref{appendix:TreesImpossibility}})  that for
any such graph, there exists a coalition  of size at most $k$ (which
is mapped to a single vertex in the simulating tree) that can bias the outcome.
\begin{restatable}{theorem}{RestatableSimulatedTreeGraphLeaderElectionImpossibility}
    \label{TreesImpossibility:LeaderElectionImpossibility}
    For every $k$-simulated tree, there is no \kresilient FLE protocol for every $\epsilon \leq \frac{1}{n}$.
\end{restatable}

\mySection{Fair Leader Election and Fair Coin-Toss are Equivalent}
\label{section:ProblemsRelations}

We reduce fair coin-toss to fair leader election by electing 
a leader and taking the lowest bit as a result. In the other
direction, we reduce leader election to fair coin-toss by tossing
$\log_2(n)$ independent coins.

Define fair coin toss similarly to FLE. A \textbf{fair coin-toss}
protocol $P$, is a symmetric protocol such that for every
oblivious schedule:
	\[\forall b\in \{0,1\}: Pr(outcome(e)=b)=\frac{1}{2}\]
where the probability is over the local randomization of the processors.

In this \sectionText we assume for simplicity that the number of processors is 
a power of $2$, i.e., $\log_2(n)$ is integer.

For simplicity, we consider only the notion of unbias (whether a protocol is 
\kunbiased) and 
not resilience, as seen previously in Lemma \ref{Model:UnbiasResilienceEquiv},
they are almost equivalent.
\begin{theorem}
	One can implement a $(\frac{1}{2}n\epsilon)$-$k$-\textit{unbiased} 
	coin-toss protocol using a \kunbiased FLE protocol.
	One can implement a $(\frac{1}{2}+\epsilon)^{n}$-$k$-\textit{unbiased} FLE 
	protocol using $log_2(n)$ independent instances of a \kunbiased protocol.
\end{theorem}
\begin{proof}
\begin{itemize}
    \item \textbf{Leader Election to Coin-Toss}:
    Run leader election to get a leader index $i\in [n]$, then output $i \pmod 2$.
    
    So,
    \[Pr(outcome_{coin}=0)\leq \sum_{i=1;i\pmod 2=0}^{n}(\frac{1}{n}+\epsilon) = \frac{1}{2}+\frac{1}{2}n\epsilon\]
    \item \textbf{Coin-Toss to Leader Election}:
    Run Coin-Toss $\log_2(n)$ times, concatenate the results and elect the processor with that index.
    
    So,
    \[Pr(outcome_{leader-election}=0) = Pr(\forall i=1...n: outcome_{coin}^i = 0) \leq (\frac{1}{2} + \epsilon)^{\log(n)}\]
\end{itemize}
\end{proof}
Notice that we assume a debatable assumption, we assume that one can execute a 
coin toss protocol $log_2(n)$ times \textbf{independently}. Without this 
assumption it is not straight-forward to induce the existence of a resilient 
FLE protocol from the existence of a resilient coin-toss protocol.

\newpage
\appendix
\mySection*{Appendix}

\mySection{Pseudo Code for \RingOriginalProtocolTitle }
\label{section:OriginalProtocolPseudoCode}

For completeness we include a Pseudo-Code for \RingOriginalProtocol,
since previous works only described it verbally.

\begin{algorithm}[H]
\caption{\RingOriginalProtocol. Resilient Leader Election on a Ring.
Code for the $origin$ processor (processor $1$).} \Fn{Init()}{
    $d_1 = Uniform([n])$\;
    Send $d_1$\;
    $round=1$\;
    $sum=0$\;
} \Fn{UponRecieveMessage($value$)}{
    $value = value\pmod n$\;
    Send $value$\;
    $round\plusplus$\;
    $sum = (sum + value)\pmod n$\;
    \If{$round==n$}{
        \uIf{$value == d_i$}{
            $Terminate(output=sum])$\;
        }
        \Else{
            $Terminate(\bot)$\;
        }
    }
}

\end{algorithm}
\newpage
\begin{algorithm}[H]
    \caption{\RingOriginalProtocol. Resilient Leader Election on a Ring. Code for $ normal $ processor $i$.}
    \Fn{Init()}{
        $d_i = Uniform([n])$\;
        $buffer = d_i $\;
        \tcp{start at 0}
        $round=0$\;
        $sum=0$\;
    }
    \Fn{UponRecieveMessage($value$)}{
        $value = value\pmod n$\;
        Send $buffer$\;\label{alg:sendValue1}
        $round\plusplus$\;
        \tcp{Put the incoming message in the buffer.}
        $buffer = value$\;\label{alg:bufferValue1}
        $sum = (sum + value)\pmod n$\;
        \If{$round==n$}{
            \uIf{$value == d_i$}{\label{alg:validateLastValue2}
                $Terminate(output=sum)$\;
            }
            \Else{
                \tcp{Validation failed.}
                $Terminate(\bot)$\;
            }
        }
    }
\end{algorithm}

\textbf{Remarks:}
\begin{itemize}[noitemsep,topsep=0pt]
    \item Swapping lines \ref{alg:sendValue1} and \ref{alg:bufferValue1} in \textit{UponRecieveMessage()} of a normal processor results in \textit{UponRecieveMessage()} of the origin.
    \item $origin$ behaves like a pipe, and $normal$ processors behave like a buffer of size $1$
    \item An execution of the protocol, can be viewed as $n$ ``rounds''. In each round, a message is sent by processor $1$ (the $origin$), then by processor $2$, etc. until a message is sent by processor $n$. Then $origin$ receives a message and initiates the following round.
\end{itemize}

\mySection{Basic, Non-Resilient Fair Leader Election Protocol}
\label{section:BasicNonResilientProtocol}

In order to clarify \RingOriginalProtocol, we present a
simpler, non-resilient FLE protocol for a unidirectional
asynchronous ring.

Each processor $i$ selects randomly uniformly a value $d_i\in [n]$
and shares it with all the other processors. Finally, the elected
leader is $\sum_{i \in [n]} d_i \pmod n$.

\begin{algorithm}[H]
    \caption{\texttt{Basic-LEAD} Basic Leader Election on a Ring. Code for processor $ i $.}
    \Fn{Init()}{
        $d_i = Uniform([n])$\;
        Send $d_i$\;
        $round=1$\;
        $sum=0$\;
    }
    \Fn{UponRecieveMessage($value$)}{
        Send $value\pmod n$\;
        $round\plusplus$\;
        \tcp{Sum all incoming messages}
        $sum = (sum + value)\pmod n$\;
        \tcp{Check whether this is the last round}
        \If{$round==n$}{
            \uIf{$value == d_i$}{\label{alg:validateLastValue}
                $Terminate(output=sum])$\;
            }
            \Else{
                \tcp{Validation failed.}
                $Terminate(\bot)$\;
            }
        }
    }
\end{algorithm}

Each processor validates that it receives $n$ values, validates that
the last value that it has received is the value that it originally
selected randomly (line \ref{alg:validateLastValue}). If not, it
aborts because some processor deviated.

\begin{claim}\label{SingleProcessorAttack}
    \texttt{Basic-LEAD} is not \kunbiased against a single adversary ($k=1$) for every $ 0 \leq \epsilon < 1-\frac{1}{n}$.
\end{claim}
\begin{proof}
    We show that processor $j$ can enforce the election of processor $w$:

    Processor $j$ might wait to receive $n-1$ incoming messages, then ``select'' its own value to cancel out their sum:
        \[d_{j} := w - \sum_{i\ne j}{d_{i}} \pmod n\]
    Then $j$ continues with the protocol execution. Thus $Pr(outcome=w) = 1 = \frac{1}{n}+(1-\frac{1}{n})$ as required.
\end{proof}

\mySection{Attacking \RingOriginalProtocolTitle with Randomly Located 
Adversaries}
\label{section:RandomizedModel}

The simple attack on \RingOriginalProtocolSpaced introduced in Lemma
\ref{Attacks:NaiveAttackLemma} can be modified to work also for
$\Theta(\sqrt{n \log(n))})$ randomly located adversaries, although
it seems possible only due to specific locations of the processors.
Moreover, we do not need to assume that adversaries know their
relative locations, i.e., $(l_i)_{i=1}^k$, and their amount $k$. In
this section, we define the randomized model and present an
appropriate attack with $\Theta(\sqrt{n \log(n))})$ adversaries. The
Cubic Attack however, relies on honest segments of increasing length
and therefore it is not adaptable to randomly located
adversaries.\medskip

\noindent\fbox{\begin{minipage}{\textwidth}
    \textbf{The Randomized Model}
        Each processor is selected to be an adversary with probability $p$, independently of the others.\\
        The adversaries do not know their relative locations $(l_i)_i$ or their exact amount $k$.
\end{minipage}}

\medskip In our attack, with good probability over selecting the adversaries and with good probability over the secret values of the honest processors, the adversaries control the outcome. We will choose $p=\Theta(\sqrt{\frac{\log(n)}{n}})$, so the expected amount of adversaries is $E[k]=\Theta(\sqrt{n \log(n))})$.

In order to adjust the naive attack from the proof of Lemma
\ref{Attacks:NaiveAttackLemma} to the randomized model defined
above, we handle two issues:
\begin{itemize}[]
    \item In the original attack each honest segment $I_j$ should be of length $l_j \leq k-1$. In order for this requirement to hold with good probability, for every segment in the randomized model, we selected $p$ to be large enough.
    \item In the previous attack the adversaries need to know their exact amount $k$ and their relative locations. To handle this issue, each adversary $a_i$ hopes for a bound over $k$ to hold and a bound over $l_i$ to hold - if these bounds hold, the attacker can guess $k$ according to its incoming messages and the attack succeeds.
\end{itemize}

\begin{theorem}
    Let $C>1$.
    There exists a symmetric adversarial deviation from \RingOriginalProtocolSpaced  such that with probability $1-\delta$ over selecting adversaries, such that $Pr(outcome=w) \geq 1 - n^{2-C}$.
    For $p = \sqrt{8\frac{\log(n)}{n}}$, $\delta = exp(-\frac{1}{2}n) + n^{-7}$.
\end{theorem}

Like in previous attacks, the attack is composed of 3 steps.

In the first step, the adversaries learn the secret values by
forwarding incoming messages. It turns out that we can learn $k$ by
observing the incoming messages. Due to circularity, after receiving
$n - k + C$ messages, we expect the last $C$ incoming messages to be
identical to the first $C$ incoming messages.

In the second step, perform calculation and a message that cancels
out the total sum of all the outgoing messages, those sent in step 1
and those that will be sent in step 3.

Last, in the third step, each adversary $a_i$ sends more $k-C-1$
messages such that the last $l_i$ messages are sent correctly. In
this step, $a_i$ hopes that $l_i \leq k-1-C$, and simply sends the
last $k-C-1$ out of the first $n-k$ first incoming messages.

\begin{proof}
    It is enough to show the adversaries can comply to condition 2 of Lemma \ref{Basic:ValidExecution} while controlling the outgoing sum of each adversary. Denote with $m[j]$ the $j^{th}$ incoming message.

    If $origin$ is selected to be adversary, let it execute honestly. \medskip

    \noindent\framebox{\parbox{\dimexpr\linewidth-1\fboxsep-2\fboxrule}{
        \textbf{Strategy for adversary $a_i$}
        \begin{enumerate}[itemsep=0pt,topsep=2pt]
            \item \label{cheat2:phaseLearn} Forward every incoming message, until we encounter circularity, i.e., the first $T>C$ such that: $m[1],m[2]..m[C]=m[T-C+1],m[T-C+2], ..., m[T]$. Calculate the estimated value of $k$, $k' := n - T + C$.
            \item \label{cheat2:phaseCancel} Send the message $M = w - S(1,T) - S((n - k') - (k' - C - 1) + 1, n - k') \pmod n$, for every $q,r$, $S(q,r):=\sum_{j=q}^{r}m[j]$ denotes a partial sum of incoming messages.
            \item \label{cheat2:sendLastSegment} Send the messages $m[(n - k') - (k' - C - 1) + 1],..., m[n - k']$.
        \end{enumerate}
    }} \medskip

    First, we show that with good probability over selecting the secret values, all adversaries finish step \ref{cheat2:phaseLearn} with $T=n-k+C$, i.e., $k'=k$.

    For a single adversary $a_i$, the series $m[1], ..., m[C]$, does not repeat twice in the series of all secret values with probability $\geq 1 - n\cdot n^{-C}$ by union bound over all possible offsets. Therefore, applying union bound over all $k \leq n$ adversaries, we get that with probability at least $1-n^{2-C}$ there is no such repetition for any adversary.

    If there is no such repetition for every adversary, then every adversary $a_i$ keeps performing step \ref{cheat2:phaseLearn} until it completes a whole cycle (sends the secret value of every honest processor) and then sends the messages $m[1], ..., m[C]$ again. At that time (just before it sends $m[1]$ for the second time), $a_i$ sent each secret once, meaning a total of $n-k$ messages. Therefore we have $T=n-k+C$ for every adversary.

    Second, we show that with high probability, $\forall j: l_j \leq k - C - 1$.

    By Hoeffding's inequality, $Pr(k \geq \frac{1}{2}np) \geq 1-exp(-\frac{1}{2}n))$. Divide the ring into $\frac{n}{\frac{1}{8}np}=\frac{8}{p}$ disjoint segments of length $\frac{1}{8}pn$. In each such segment, there is an adversary with good probability, which implies that the maximal distance between two adversaries is at most $2\frac{1}{8}pn=\frac{1}{4}pn$. Which is smaller than $k-C-1$ with good probability.
    \[Pr(\textit{a segment of length }\frac{1}{8}pn\textit{ does not contain an adversary}) = (1-p)^{\frac{1}{8}pn} =\]
    \[= {(1-p)^{\frac{1}{p}}}^{\frac{1}{8}np^2} \approx exp(-\frac{1}{8}n p^2) = exp(-8\log(n)) = n^{-8}\]
    Applying the union bound over all the $\frac{8}{p} < n$ segments described above, gives
        \[Pr(\forall j: l_j \leq k-C) \geq 1 - exp(-\frac{1}{2}n)-n\cdot n^{-8} = 1 - \delta\]

    Third, notice that due to step \ref{cheat2:phaseCancel} and the definition of the outgoing messages of $a_i$, the total sum of its outgoing messages is
        \[S(1,T) + M + S((n - k') - (k' - C - 1) + 1,n-k') = w \pmod n \]
    If previously required conditions hold ($k-C-1\geq l_j$ and $T=n-k+C$) for every adversary, then the last $l_j \leq k-C-1=k'-C-1$ outgoing messages of $a_j$ are the secret values of the honest processors in $I_j$.

    Last, notice that the if the conditions hold then total number of outgoing messages of $a_i$ is $n = T + 1 + (k - C - 1)$ as required by condition 1 of Lemma \ref{Basic:ValidExecution}.

    Therefore, with high probability all conditions of Lemma \ref{Basic:ValidExecution} hold, as required.
\end{proof}

\chapter{Pseudo-Code for the Cubic Attack}
\label{section:CubicAttackPseudoCode}

In this \sectionText we outline an explicit pseudo-code for the Cubic Attack 
outlined in \SectionText \ref{Attacks:CubicAttackTheorem}.

\begin{algorithm}[H]
    \caption{\textbf{CubicAttack}
        Code for adversary $a_i$, elected leader is $w$.}
    \Fn{Init()}{
        Init array $m[1...n-k]$\;
        $count = 0$\;
    }
    \Fn{UponRecieveMessage($v$)}{
        $count\plusplus$\;
        $m[count] = v$\;
        \If{$count \leq n-k-l_i$}{
            Send $v$\;
        }
        \If{$count == n-k-l_i$}{
            \For{$i = 1$ \KwTo $k-1$}{
                Send $0$\;
            }
        }
        \If{$count==n-k$}{
            $S = \sum_{j=1}^{n-k} m[j]$\;
            Send $w - S$\;
            \For{$j = n-k-l_i+1$ \KwTo $n-k$}{
                Send $m[j]$\;
            }
            $Terminate(output=w)$\;
        }
    }
\end{algorithm}\medskip

\mySection{\RingOriginalProtocolTitle is 
\texorpdfstring{$\epsilon$-$\sqrt[4]{n}$-resilient}{epsilon-sqrt[4]\{n\}-resilient}}
\label{section:ALeadIsResilient}

It is enough to consider only deterministic deviations, because if
there exists a probabilistic adversarial deviation that obtains
$Pr(outcome=j) > \frac{1}{n}+\epsilon$ then there exists a
deterministic adversarial deviation with $Pr(outcome=j) >
\frac{1}{n}+\epsilon$.

Recall the message schedule has no impact on the local calculations
of every processor, so given a deviation, the execution is
determined by the randomization of the honest processors,
$\DataValues$. Call these values the \textbf{input}. So an event in
the probability space is a subset of the input space $\chi :=
[n]^{n-k}$.

Let $NoFail^D \subseteq \chi$ be the event that the execution
completes with a valid output, i.e., all processors terminate with
the same $output\ne \bot$. We upper bound $Pr(outcome=j)$ for $j=1$,
the same analysis holds for any other $j$. Recall that adversaries
can always reduce this probability to zero simply by aborting
whenever the output should be $j$.

\begin{claim}
    \label{thm:LinearResilienceRing}
For every consecutive coalition along the ring $C\subseteq
\{1,\ldots , n\}$ of size $k$,  for every $j\in [n]$, for every
adversarial deviation from \RingOriginalProtocolSpaced with adversaries
$C$: $Pr(outcome=j) \leq
    \frac{1}{n}$.
\end{claim}
\begin{proof}
    Let $C$ be a continuous coalition of size $k$. For simplicity, assume the $origin$ is an honest processor. There is a single honest segment of length $k>l$, denote it with $I$. Denote the only exposed adversary with $a$. Denote the successor of $a$ with $h$, it is honest because $a$ is exposed. Let $j\in[n]$. Let $P$ be some deterministic adversarial deviation from \RingOriginalProtocolSpaced with the coalition $C$. Observe the probability space over the randomizations of the honest processors. Recall that given the randomizations, there is a single corresponding execution of the deviation.

    For all $i$, let $r_i$ be the $i^{th}$ outgoing message of $a$.
    Denote $Sum_b = \sum_{i=1}^{n}{r_i} \pmod n$
    Denote the random values of $I$ with $d_b, d_{b+1}, ..., d_{b+l-1}$.
    From Lemma \ref{Basic:ValidExecution}, we have
    \[Pr(outcome=j) = Pr(Sum_b=j \wedge NoFail^D)\]
    More, from Lemma \ref{Basic:ValidExecution}, we have
    \[Pr(Sum_b=j \wedge NoFail^D) = Pr(Sum_b=j \wedge \forall i=0,...{l-1}: d_{b+i}=r_{n-i}) \leq \]
    \[\leq Pr(Sum_b = \sum_{i=1}^{n-l}{r_i} +\sum_{i=0}^{l-1}{d_{b+i}}))\]
    Define a sum $S$ such that $\sum_{i=1}^{n-l}{r_i} +\sum_{i=0}^{l-1}{d_{b+i}} = S+d_b$.

    Upon receiving $k-1$ messages, the honest segment sends at most $k\leq l-1$ messages. The first $l-1$ outgoing messages of $I$ are just $d_{b+1}, ..., d_{b+l-1}$, which are independent with $d_b$. Therefore the first $k=n-l$ outgoing messages of $a$ are independent with $d_b$. So the sum $S$ is independent with $d_b$. Since $d_b$ distributes uniformly in $[n]$, also $S+d_b$ distributes uniformly.

    So, $Pr(outcome=j) \leq Pr(S+d_b=j) = \frac{1}{n}$ as required.
\end{proof}

\subsomething{Detailed proof of \texorpdfstring{$\sqrt[4]{n}$}{sqrt[4]{n}}-resilience for
\RingOriginalProtocolTitle}
 Since we focus only on deterministic
deviations, consecutive adversaries ($a_i, a_{i+1}$ s.t. $l_i=0$) do not need to
communicate anything more than the incoming messages of $a_i$. That
is, one may assume that $a_i$ always behaves like a pipe i.e., it
transfers every incoming message to $a_{i+1}$ and does not send any
other message besides that. For that reason, the analysis focuses on
exposed adversaries ($a_i$ such that $l_i \geq 1$).

It is enough to consider deterministic (non-probabilistic)
adversarial strategies. If there exists a probabilistic adversarial
deviation that obtains $Pr(outcome=j) > \frac{1}{n}+\epsilon$ then
there exists a deterministic adversarial deviation with
$Pr(outcome=j) > \frac{1}{n}+\epsilon$. For a probabilistic
deviation, the success probability $Pr(outcome=j)$ is the average
success probability over the various results of randomizations of
the deviation. Therefore, there exists a result for the
randomizations for which $Pr(outcome=j)$ is larger or equal than its
expectancy, as required.

Given an adversarial deviation $D$, for every input $x\in \chi$,
denote by $out_i(x)$ the list of outgoing messages of $a_i$ along
the execution of $x$.

\begin{lemma}
    \label{Quad:OutIsInjective}
For every adversary $a_i$, and for every $x,x'\in NoFail^D$ such
that $out_i^D(x)=out_i^D(x')$, then $x=x'$.
\end{lemma}
\begin{proof}
    The last $l_i$ outgoing messages of $a_i$ in $x$ and in $x'$ are identical, therefore by Lemma \ref{Basic:ValidExecution}, the secret values of $I_i$ are identical in $x$ and in $x'$. But the incoming messages of $a_{i+1}$ are the secret values of $I_i$ and some of the outgoing messages of $a_i$, therefore the incoming messages of $a_{i+1}$ are identical in $x$ and in $x'$. The deviation is deterministic, therefore $out_{i+1}(x)=out_{i+1}(x')$. Continue by induction on $i$, and we get that $out_j(x)=out_j(x')$ for all $j$ and therefore the secret values of $I_j$ are identical in $x$ and in $x'$. Therefore $x=x'$ as required.
\end{proof}

For a time $t$, denote by $Sent_i^t$ ($Recv_i^t)$ the amount of
messages sent (received) by $a_i$ until, and including, time $t$.

Recall $k_0=\frac{1}{4}\sqrt[4]{n}$.

\begin{restatable}{lemma}{RestatableQuadSendLikeRecv}
    \label{Quad:SendLikeRecv}
    For every adversarial deviation $D$
    \[Pr(NoFail^D \wedge \exists t, i: Sent_i^t > Recv_i^t+2k) \leq n^{2-k_0}\]
\end{restatable}
\begin{proof}
    For every $a_i$ and $r< n$, define the following set of bad inputs
    \[ B_{r,i} := \{x \in \chi ~|~ \exists t \textup{ s.t. }: Recv_i^t = r, ~~Sent_t^i > r + 2k\} \]
    So we have $| \{out_i(x) ~|~ x\in B_{r,i} \} | \leq n^r n^{n-2k-r} $ because there are at most $n^r$ options for the first $r+2k$ messages from the definition of $B_{r,i}$, and at most $n^{n-2k-r}$ options for the later $n-2k-r$ messages.\\
    Since, $n^r n^{n-2k-r} = \frac{|\chi|}{n^k}$ and that $out_i$ is injective on non-failing inputs by Lemma \ref{Quad:OutIsInjective}, we get that $Pr(NoFail^D \cap B_{r, i}) \leq n^{-k}$.\\
    Apply the union bound over all $nk < n^2$ options for $r$ and $i$ and obtain
    \[Pr(NoFail^D \wedge \exists t, i: Sent_i^t > Recv_i^t+2k) \leq n^2n^{-k} = n^{2-\frac{1}{4}\sqrt{n}}\]
\end{proof}

\begin{lemma}
    \label{Quad:RecvLimit}
    For every adversarial deviation, for every time $t$, for every $i \in [k]$, $Recv_{i+1}^t \leq Sent_i^t$
\end{lemma}
\begin{proof}
    At initialization, all of the honest processors are idle (recall the $origin$ is adversary), and each honest processor responds to an incoming message with a single outgoing message. More, at initialization we have $Recv_{i+1}^{0} = 0 = Sent_i^{0}$ for every $i \in [k]$. So if for some later time $t$, $Recv_{i+1}^{t_0} = r$, then the (honest) predecessor of $a_i$ sent $r$ messages by time $t$, therefore its predecessor sent $r$ messages by time $t$ etc. Therefore, $a_{i-1}$ sent $r$ messages by time $t$. So $Recv_{i+1}^t = r \leq Sent_i^t$.
\end{proof}

\begin{restatable}{lemma}{RestatableQuadSquareDiff}
    \label{Quad:SquareDiff}
    For every adversarial deviation $D$
    \[Pr(NoFail^D \wedge \exists t, i, j: |Sent_i^t - Sent_j^t | > 2k^2) \leq n^{2-k_0}\]
\end{restatable}
\begin{proof}
    By Lemma \ref{Quad:SendLikeRecv}, it enough to show that for every input $x$ such that $\exists t, i, j: |Sent_i^t - Sent_j^t | > 2k^2$, there exists an adversary $i'$ such that $Sent_{i'}^t > Recv_{i'}^t + 2k$.\\
    We show it by applying the opposite inequality along the ring and applying Lemma \ref{Quad:RecvLimit}. Let $t$ be a time, $a_i,a_j$ be two adversaries. Assume that for all $i'$: $Sent_{i'}^t \leq Recv_{i'}^t + 2k$. In particular $Sent_i^t \leq Recv_i^t + 2k$ and by Lemma \ref{Quad:RecvLimit}, $Recv_i^t \leq Sent_{i-1}^t$ - so $Sent_i^t \leq Sent_{i-1}^t + 2k \leq ... \leq Sent_j^t + 2k(i-j \pmod k) \leq Sent_j^t + 2k^2$. By Symmetry, $Sent_j^t \leq Sent_i^t + 2k^2$. Therefore $|Sent_i^t - Sent_j^t | \leq 2k^2$.
\end{proof}

\RestatableMainUnbiasTheorem*
\begin{proof}
    Denote the ``$k^2$-synchronized'' inputs of $D$ as $Sync^D = \{x\in \chi ~|~ \forall i,j,t: |Sent_i^t -Sent_j^t| \leq 2k^2\}$.

    By the law of total probability, $Pr(outcome=1) \leq Pr(NoFail^D \wedge x \notin Sync^D) + Pr(outcome=1 \wedge x \in Sync^D)$. By Lemma \ref{Quad:SquareDiff}, $Pr(outcome=1 \wedge x\notin Sync^D) \leq Pr(NoFail^D \wedge x \notin Sync^D) \leq \epsilon$. Therefore, it is enough to show that $Pr(outcome=1 \wedge x\in Sync^D) \leq \frac{1}{n}$.

    Let $h_1$ be the honest successor of $a_1$. Let $q_{h_1}(x) \subseteq \chi$ be the set of $n-1$ inputs that differ from $x$ only in the secret value of $h_1$, $d_{h_1}$. It is enough to show that for every $x\in Sync^D $ such that $outcome=1$, we have for every $x' \in q_{h_1}(x)\cap NoFail^D\cap Sync^D$: $outcome \ne 1$.

    Let $x\in Sync^D$ such that $outcome=1$, let $x' \in q_{h_1}(x)\cap NoFail^D \cap Sync^D$. By Lemma \ref{Basic:ValidExecution} the last $l_1$ outgoing messages of $x$ and $x'$ differ only in $d_{h_1}$, therefore it is enough to show that the first $n-l_1$ outgoing messages of $a_1$ are independent of $d_{h_1}$ and therefore identical in $x$ and in $x'$ because then the sum of outgoing messages of $a_1$ differs in the two executions.\\

    Observe the execution of $x$. For every adversary $a_i$, denote by $t_i$ the first time that $ a_i $ receives a message that might depend on $ d_{h_1} $. Intuitively, $t_ i$ is the first time that information about $d_{h_1}$ reaches $a_i$. As explained in the previous paragraph, it is enough to prove that $Sent_1^{t_1} \geq n-l_1$.

    For $k \geq i > 1$, the adversary $a_{i+1 \pmod k}$ receives a message that depends on $d_{h_1}$ only after $a_i$ sends such a message and sends further $ l_i $ messages afterwards (to “push” it along $ I_i $). Therefore, $ a_i $ sends at least $ l_i+1 $ messages between $ t_i $ and $ t_i+1 $. Which implies
    $l_i \leq  l_i + 1 \leq Sent_i^{t_{i+1}}-Sent_i^{t_i} (*)$.\\

    \textit{Claim}: For every $ k+1 \geq i \geq 2 $, $ Sent_1^{t_i} \geq l_1+l_2+... + l_{i-1} - 4(i-2)k^2 $\\
    (where the claim for $i=k+1$ states $Sent_1^{t_1} \geq l_1+l_2+... + l_{k} - 4(k-1)k^2$)\\
    \textit{Proof by induction on i}:
    \begin{itemize}[noitemsep,topsep=0pt]
        \item \textbf{Basis}: $ i=2 $, only after $ a_1 $ sends $ l_1 $ messages, $ a_2 $ receives the message $d_{h_1}$. Therefore $ Sent_1^{t_2} \geq l_1 + 0$ as required.
        \item \textbf{Step}: Assume for $ i $, show for $ i+1 $. From (*) we have
        $ l_i \leq Sent_i^{t_{i+1}} - Sent_i^{t_i} $
        Therefore, since $x, x' \in Sync^D$, we get $l_i - 4k^2 \leq Sent_1^{t_{i+1}} - Sent_1^{t_i}$, then plugging it into the induction hypothesis we obtain the wanted inequality
        \[ Sent_1^{t_{i+1}} \geq Sent_1^{t_i} + l_i - 4 k^2 \geq l_1+l_2+... + l_{i-1} - 4(i-2)k^2 + l_i - 4k^2 = l_1+l_2+... + l_{i-1} + l_i - 4(i+1-2)k^2 \]
    \end{itemize}
	\medskip
    Substitute $i=k+1=1\pmod k$ in the claim above and obtain,
    \[Sent_1^{t_1} \geq l_1 + l_2 +... + l_k - 4(k-1)k^2 \geq n-k-4k^3\]
    However, we know that $l_1 \geq 60k^3 > k+4k^3$, therefore $Sent_1^{t_1} \geq n-l_1$.
    
\end{proof}

\mySection{\PhaseAsyncLeadTitle is 
\texorpdfstring{$\epsilon$-$\sqrt{n}$-resilient}{epsilon-sqrt\{n\}-resilient}}
\label{section:PhaseAsyncLeadFull}

In this section, we prove that \PhaseAsyncLead
is $O(\sqrt{n})$-resilient. For completeness, we include a verbose pseudo-code
for this protocol in
\SubSomethingText \ref{Sqrt:Pseudo-Code}.

\subsomething{Preliminaries and notation for a full resilience proof}

Let $Honest := V \backslash C$ be the set of honest processors.

Let $NoFail^D \subseteq \chi$ be the event that the execution completes with a valid output, i.e., all processors terminate with the same output, $o\ne \bot$. We bound the probability of the event $outcome=j$ for every $j$, i.e., the protocol completes and the output of all the honest processors is $j$. The function $f$ is random, so w.l.o.g we focus on $j=1$, and upper bound $Pr(outcome=1)$.\\

Most of the proof focuses on showing that the synchronization
mechanism indeed keeps all processors synchronized, i.e.
$Pr(NoFail^D ~|~ \chi_1^D) \leq \frac{1}{n}$.

Since we focus only on deterministic deviations, consecutive
adversaries ($a_{i+1}=a_i+1$) do not need to communicate anything
more than the incoming messages of $a_i$. That is, one may assume
that $a_i$ always behaves like a pipe i.e., it transfers every
incoming message to $a_{i+1}$ and does not send any other message
besides that. For that reason, the analysis focuses on exposed
adversaries.

In every execution we have two types of events - The first, $send(p, i)$ is the event that the processor $p$ sends its $i^{th}$ outgoing message. The second, $recv(p, i)$ is the event that $p$ receives its $i^{th}$ incoming message. Notice that the $i^{th}$ outgoing ``validation'' message of $p$ is its $2i^{th}$ outgoing message. Denote the set of all events as $Events=\{send(p, i) ~|~ p \textup{ is a processor}, 1 \leq i \leq 2n \} \cup \{recv(p, i) ~|~ p \textup{ is a processor}, 1 \leq i \leq 2n \}$. Notice that this set is independent of the deviation or the input. Note that for each message there are two correlating events - one for its dispatch ($ send $) and the other for its arrival and processing ($recv$).\\

As a shorthand, denote $s(p) = send(p, 2p)$ - the event of $p$ sending its validation message as a validator. Similarly, denote $r(p) = send(p - 1, 2p)$ - the event of $p-1$, the predecessor of $p$, sending the validation message of $p$. Note that there is a significant difference between the event $r(p) =  send(p-1, 2p)$ and the receiving event $recv(p, 2p)$, which is triggered by $r(p)$. For example, the event $s(p)$ always happens before $send(p, 2p+1)$, but the event $r(p)=send(p-1, 2p)$ might happen before $s(p)=send(p, 2p)$.\\

Some of the events never occur in some anomalous executions (i.e., if the protocol runs forever). Given a deviation $D$ and an input $x$ are clearly determined by the context, if an event $e\in Events$ does occur in an execution, then define its \textbf{data}, $data(e)$, as the content of the actual message sent/processed by the relevant processor.\\

Next, we define two directed graphs on the set of all possible events, $Events$. The graphs represent the dependencies of messages sent by processors in an execution of a \textbf{specific deviation} $D$ on a \textbf{specific input} $x$.\\

The first graph is the ``happens-before'' graph. It is denoted with $G^D_x$, where $D$ is a deviation and $x\in \chi$ is an input. Given a deviation $D$ is determined clearly by the context, for two events $\alpha, \beta\in Events$, the notation $\alpha \to_x \beta$ (or simply $\alpha \to \beta$ when $x$ is determined clearly by the context) means the pair $(\alpha, \beta)$ is an edge in the graph $G^D_x$.\\
Intuitively, in this graph there exists a route from an event
$\alpha$ to another event $\beta$ if and only if the event $\alpha$
happens before the event $\beta$ for \textbf{every message schedule}
(that is, for every execution). For completeness, the edges of
$G^D_x$ are defined explicitly below:
\begin{itemize}
    \item Arrival edges: A message is received by processor $p + 1$ after it was sent by $p$, $send(p, i) \to_x recv(p + 1, i)$.
    \item Local linearity edges: The $i^{th}$ message is always sent/processed before the $(i+1)^{th}$ message, so we define $ send(p, i) \to_x send(p, i+1), recv(p, i) \to_x recv(p, i + 1)$.
    \item Triggering edges:  If the $ i^{th} $ incoming message of $p$ \textbf{triggered} the delivery of its $j^{th}$ outgoing message, then $recv(p, i) \to_x send(p, j)$. That is, $p$ sends its $j^{th}$ outgoing message while processing its $i^{th}$ incoming message.
    \item Receive after sending: If $p$ expects the $i^{th}$ incoming message only after sending the $j^{th}$ outgoing message then $send(p, j) \to_x recv(p, i)$. More formally, this edge exists if $send(p, j)$ was triggered by an event prior to $recv(p, i)$, that is: $\exists i' < i~~s.t.: recv(p, i') \to_x send(p, j)$.
\end{itemize}

The second graph is the ``calculations-dependency'' graph. It is denoted with $Gc^D_x$, where $D$ is a deviation, $x\in \chi$ is an input. Given a deviation $D$ is clearly determined by the context, for two events $\alpha, \beta\in Events$, the notation $\alpha \to_{c, x} \beta$ (or simply $\alpha \to_{c} \beta$ when $x$ is determined clearly by the context) means the pair $(\alpha, \beta)$ is an edge in the graph $Gc^D_x$.\\
Intuitively, there exists a route from an event $\alpha$ to another
event $\beta$ if the calculation of $data(\beta)$ depends on
$data(\alpha)$. For completeness, the edges of $Gc^D_x$ are defined
below:
\begin{itemize}
    \item Send to Receive: The data of a message depends on itself, so $send(p, i) \to_{c,x} recv(p+1, i)$.
    \item Validation value immediate transfer: When an honest processor $h$ receives a validation message, it forwards its content immediately - $recv(h, 2i) \to_{c,x} send(h, 2i)$ (For every $h \ne i$).
    \item Data values delay: When an honest processor $h$ receives a data message, it forwards its content only on the following round - $recv(h, 2i-1) \to_{c,x} send(h, 2i+1)$ (For every $1 < i < n$).
    \item General calculation of adversaries: When the event $recv(a_i, j_1)$ triggers $send(a_i, j_2)$, the adversary $a_i$ processes the message of $recv(a_i, j_1)$ and calculates a messages list to send (it sends $0$ to $n$ messages) - for a general deviation, $data(send(a_i, j_2))$ might depend on all its preceding incoming messages, that is: $\forall t\leq j1: recv(a_i, t) \to_{c, x} send(a_i, j_2)$.
\end{itemize}

Given a deviation $D$ and an input $x$, denote $\alpha \leadsto \beta$ if there is a path from $\alpha$ to $\beta$ in the happens-before graph $G^D_x$ and denote $\alpha \leadsto_c \beta$ if there is a path from $\alpha$ to $\beta$ in the calculations graph $Gc^D_x$. Denote $\alpha \not\leadsto \beta$, and $\alpha \not\leadsto_c \beta$ correspondingly if such paths do not exist.\\

\begin{remark} Calculation dependence is stronger than ``happens-before'' relation. That is, if $\alpha \leadsto_{c,x} \beta$, then also $\alpha \leadsto_x \beta$, but not necessarily the other way around.
\end{remark}
\begin{remark}Both $Gc^D_x$ and $G^D_x$ are cycle free.
\end{remark}
Denote the \textit{point of no return} as the event $nr=send(a_1, 2(n-l)) \in Events$. This is the point of no return of the adversary $a_1$. After that point $a_1$ already sent all the messages that affect the output calculated by honest processors in $I_1$, except for its last $l$ outgoing data messages. However, its last $l$ outgoing data messages are predetermined, because they must be the appropriate data values of the first $l$ honest processors in $I_1$, $\{d_h ~|~ h\in\{a_1, a_1+1,..., a_1+l\} \}$, or else one of them will abort.\\

The probabilities in this \sectionText are calculated over selecting an input $x\in \chi$ uniformly, given a coalition $C$ and a deviation $D$. When it is calculated over selecting a random function $f$, we denote it by $Pr_f(...)$ explicitly.\\

Denote the\textbf{ trivial deviation }as the deviation in which all adversaries behave honestly, that is they execute the code of the protocol.\\

Denote $M_0 = 2(n -l + 4k + 2)$.\\
Recall $l_i$ denotes the length of the $i^{th}$ honest segment, the segment between $a_i$ and $a_{i+1}$.\\
Given a deviation $D$ and an input $x$, let $F^D_x$ be the events that
are triggered at an early stage of the execution - before any $send(a_i, M_)$
takes place. Formally,

\[F^D_x = \{e \in Events ~|~ e\textup{ occurs }, \forall a_i \textup{ s.t. }
l_i\geq 1: send(a_i, M_0) \not\leadsto_x e \}\]

For every input, considering the trivial deviation (each adversary
behaves honestly), the validation processes occur one after the
other without intersection, that is $s(1) \leadsto r(1) \leadsto
s(2) \leadsto r(2) \leadsto ... \leadsto r(n)$. Breaking this order
can be thought of as breaking the artificial synchronization implied
by the validators. For a general deviation $D$, we define the inputs
for which the processors break the synchronization significantly
before the point of no return as $Async^D$.

\[ Async^D := \{ x \in \chi ~|~ nr \notin F^D_x \} = \{x\in \chi ~|~ \exists a_i, \textup{ s.t. } l_i \geq 1,~~ send(a_i, M_0) \leadsto_x nr \} \]

In order to see that the definition above implies a significant violation of
the expected synchronization, notice that for the trivial deviation we have
$nr=send(a_1, 2(n-l)) \leadsto r(n-l) \leadsto ... \leadsto r(n-l+4k) \leadsto
send(a_i, 2(n-l+4k+2))=send(a_i, M_0)$. So intuitively, inputs in $Async^D$
violate the synchronization in at least $2\cdot 4k=8k$ messages.

Next, we prove that due to the phase validation mechanism, inputs in $Async^D$
are irrelevant.\\
The proof relies heavily on the delicate details of the phase
validation mechanism.

\begin{definition}
    Let $h\in Honest$ be an honest processor. For every input $x \in \chi$, define
    \[q_h(x) = \{x' \in \chi ~|~ x'\ne x, ~x \textup{ and } x' \textup{ differ only in the validation value of } h, ~ v_h\}\]
\end{definition}
Notice that $|q_h(x)| = m - 1$.

\subsomething{Full resilience proof for \PhaseAsyncLeadTitle}

\begin{lemma}
    \label{ChangeOneValue}
    Given a deviation and an honest processor $h$, for every event $e\in Events$, for every input $x \in \chi$, for every input $x'\in q_h(x)$, if $e$ happens in $x$ and $s(h) \not\leadsto_{c,x} e$ then (1) $s(h) \not\leadsto_{c,x'} e$ (2) $data(e)$ is identical in $x$ and in $x'$.
\end{lemma}

\begin{proof}
    Assume $s(h) \not\leadsto_{c,x} e$. Define the events required to calculate
    $e$: $Pre(e) = \{\alpha \in Events ~|~ \alpha \leadsto_{c, x} e \}$. So the
    induced graph $Gc^D_x |_{Pre} $ forms a calculation tree.\\

    We prove that for every event $e$ that happens in $x$, $Tree_x := Gc^D_x |_{Pre(e)} = Gc^D_{x'} |_{Pre(e)} =: Tree_{x'}$ by induction on the depth of $Gc^D_x |_{Pre} $.
    \begin{itemize}[noitemsep]
        \item \textbf{Basis}: The depth is zero, meaning $e$ is a root in $Tree_x$ and therefore also in $Tree_{x'}$ so they are equal.
        \item \textbf{Step}: For every predecessor $e'$ of $e$ in $Tree_x$, we have $s(h) \not\leadsto_{c, x} e'$ because $e' \leadsto_{c,x} e$. So the calculation tree of $e'$ is identical in $x$ and in $x'$. This is true for every predecessor $e'$, therefore $Tree_x = Tree_{x'}$.
    \end{itemize}
    The calculation trees are identical so in particular the predecessors of $e$ (in $Gc^D_x$ and in $Gc^D_{x'}$) are identical. The event $s(h)$ is not in the tree so all the calculations in that tree are not affected by the value of $v_h$ - therefore they are identical in $x$ and in $x'$ so $data(e)$ is also identical.
\end{proof}

\begin{definition}
    Given a deviation and an input $x$, an honest processor is \textbf{validated} in $x$ if $s(h) \leadsto_{c,x} r(h)$. Otherwise, $s(h) \not\leadsto_{c,x} r(h)$, we say it is \textbf{unvalidated}.
\end{definition}
The Lemmas below apply for every deviation $D$ for every input $x\in
\chi$.
\begin{lemma}
    \label{Sqrt:StoR}
    If $s(h) \leadsto r(h)$ then for every other honest processor $h' \ne h$, there exists an event of the form $send(h', j)$ such that $s(h) \leadsto send(h', j) \leadsto r(h)$.
\end{lemma}

\begin{proof}
    If $h'=h-1$ then take $r(h)=send(h', 2h)$ and we are done.\\
    Otherwise, recall $s(h)=send(h,2h)$, $r(h)=send(h-1,2h)$. In particular, $s(h)$ is performed by $h$ and $r(h)$ is performed by the predecessor of $h$. More, every edge in the happens-before graph connects a processor to itself or a processor to its successor. Therefore, the path from $s(h)$ to $r(h)$ must pass through $h'$. Therefore, the path must pass through some node of the form $send(h', j)$ as required.\\
\end{proof}

\begin{lemma}
    \label{Sqrt:StoRcalc}
    If $s(h) \leadsto_c r(h)$ then for every other honest processor $h' \ne h$, there exists an event of the form $send(h', 2j)$ (i.e., a validation message) such that $s(h) \leadsto_c send(h', 2j) \leadsto_c r(h)$.
\end{lemma}

The proof similar to the proof of Lemma \ref{Sqrt:StoR}, but it is
more delicate.

\begin{proof}
    Let $P$ be a path in the calculation-dependency graph from $s(h)$ to $r(h)$.\\
    Split into cases:
    \begin{itemize}[noitemsep,topsep=0pt]
        \item The processors $h', h$ are located in different segments. The path $P$ must pass through $h'$, so let $j$ be the \emph{minimal} integer such that the path passes through the node $send(h', j)$. If $j$ is even, then we are done. Otherwise, it is odd - meaning that the event $send(h', j)$ corresponds to a data message. Next, we show that there is a similar path that passes through $send(h', j-1)$ which completes the proof because $j-1$ is even.\\
        Let $a_i, a_{i-1}\in C$ be the adversaries located behind and after honest segment of $h'$. Recall the definition of the outgoing edges of $h'$ in the calculation-dependency graph. So $P$ has a sub-path of the form $send(a_{i-1}, j-2-t)\to_c... \to_c recv(h', j-2) \to_c send(h', j) \to_c ... \to_c recv(a_i, j+t') \to_c ...$ for some $t,t' \geq 0$. Let $P'$ be the following modification of $P$: Before $send(a_{i-1}, j-2-t)$ and after $recv(a_i, j+t')$ it is identical to  $P$. Then, instead of passing through the data message, it passes through the validation message using the following path $send(a_{i-1}, j-1) \to_c .. \to_c recv(h', j-1) \to_c send(h', j-1) ... \to_c recv(a_i, j-1)$. From the definition of calculation-dependency edges of adversarial processors this is also a path in the calculation-dependency graph. The path $P'$ passes through $send(h', j-1)$ as required.
        \item The processors $h'$ and $h$ are located in the same honest segment and $h'$ is located before $h$ on their segment. That is, the segment is of the form $I=\{h1, h1+1,..., h',..., h, h+1,...\}$. So the suffix of every path to $r(h)$ in the calculations-dependency graph is of the form $send(h1, 2h) \to_c recv(h1+1, 2h) \to_c send(h1+1, 2h) \to...\to_c send(h', 2h) \to...\to_c send(h-1, 2h)=r(h)$. In particular, $P$ passes through $send(h',2h)$ as required.
        \item The processor $h'$ is in the same honest segment as $h$ and it is located after $h$ on their segment. That is, the segment is of the form $I=\{h1, h1+1,..., h,..., h', h'+1,...\}$. So the prefix of every path to $r(h)$ in the calculations-dependency graph is of the form $send(h, 2h) \to_c recv(h+1, 2h) \to_c send(h+1, 2h) \to...\to_c send(h', 2h)$. In particular, $P$ passes through $send(h',2h)$ as required.
    \end{itemize}
\end{proof}

\begin{lemma}
    \label{Sqrt:SimpleLock}
    Let $h$ be an honest processor, let $s(h)\ne send(h,j) \in Events$ be an event. If $s(h) \leadsto send(h,j)$ then $r(h) \leadsto send(h,j)$. If $send(h,j) \leadsto r(h)$ then $send(h,j) \leadsto s(h)$.
\end{lemma}
\begin{proof}
    By definition of the happens-before graph $\forall t\geq 1: r(h) \leadsto send(h, 2h+t)$.\\
    Assume $s(h) \leadsto send(h,j)$, then $j > 2h$. Therefore $j=2h+t$ for some $t\geq 1$, so $r(h) \leadsto send(h,j)$ as required.\\
    Assume $send(h,j) \leadsto r(h)$. If we had $j > 2h$, then we had $r(h) \leadsto send(h, j)$, therefore $j \leq 2h$. So $send(h,j) \leadsto send(h, 2h)=s(h)$ as required.
\end{proof}

Given a deviation $D$ and an input $x$, two events $e1, e2\in
Events$ are called \textbf{simultaneous} in $x$ if $e1
\not\leadsto_x e2$ and also $e2 \not\leadsto_x e1$.

\begin{lemma}
    \label{Sqrt:Simultaneous}
    For every deviation, let $x\in \chi$.
    For every two honest processors $h1,h2$, if $s(h1) \leadsto r(h1), s(h2) \leadsto r(h2)$, then the events $s(h1),s(h2)$ are not simultaneous in $x$. In particular, if they are validated then $s(h1), s(h2)$ are not simultaneous.
\end{lemma}

\begin{proof}
    Assume by contradiction that $s(h1),s(h2)$ are simultaneous.\\
    Let $P1$ be a path from $s(h1)$ to $r(h1)$ and $P2$ be a path from $s(h2)$ to $r(h2)$. From Lemma \ref{Sqrt:StoR}, $P1$ passes through a node $e2=send(h2, j)$. The happens-before graph induces a total order over the events of $h2$, therefore either $e2 \leadsto s(h2)$ or $s(h2) \leadsto e2$. But if $e2 \leadsto s(h2)$ then we get that $s(h1)\leadsto e2\leadsto s(h2)$ which contradicts the assumption. Therefore, $s(h2) \leadsto e2$. Then by Lemma \ref{Sqrt:SimpleLock}, $r(h2) \leadsto e2$.\\
    Symmetrically, we define $e1$ and get also $s(h2) \leadsto e1 \leadsto r(h2)$, $r(h1) \leadsto e1$. Overall there is a cycle $e1 \leadsto r(h2) \leadsto e2 \leadsto r(h1) \leadsto e1$. But the happens-before graph is cycle free. Contradiction.
\end{proof}

\begin{lemma}
    \label{Sqrt:SequentialSegment}
    For every two sequential honest processors $h,h+1$:
    \begin{enumerate}[noitemsep,topsep=0pt,leftmargin=6\parindent]
        \item \label{Sqrt:SequentialSegment1} $r(h) \leadsto s(h+1)$
        \item \label{Sqrt:SequentialSegment2} $r(h) \leadsto r(h+1)$
        \item \label{Sqrt:SequentialSegment3} $s(h) \leadsto s(h+1)$
    \end{enumerate}

\end{lemma}

\begin{proof}
    By the definition of the happens-before graph,
    \begin{enumerate}[noitemsep,topsep=0pt]
        \item $r(h)=send(h-1, 2h) \to recv(h, 2h) \to send(h, 2h+1) \to recv(h+1, 2h+1) \to send(h+1, 2h+2)=s(h+1)$
        \item $r(h)=send(h-1, 2h) \to send(h-1, 2(h+1)) \to recv(h, 2(h+1)) \to send(h, 2(h+1)) = r(h+1)$
        \item $s(h)=send(h, 2h) \to recv(h+1, 2h) \to send(h+1, 2h) \to send(h+1, 2h+2)= s(h+1)$
    \end{enumerate}
\end{proof}

\begin{lemma}
    \label{Sqrt:SequentialLock}
    Let $h,h+1, h'$ be three distinct honest processors ($h,h+1$ are consecutive). If $s(h) \leadsto s(h') \leadsto_c r(h') \leadsto r(h+1)$ then $h$ is unvalidated or $h+1$ is unvalidated (possibly both).
\end{lemma}
\begin{proof}
    From Lemma \ref{Sqrt:StoRcalc}, there exists a validation message event $send(h+1, 2j)$ such that $s(h') \leadsto_c send(h+1, 2j) \leadsto_c r(h')$. Assume by contradiction that both $h,h+1$ are validated. The only validation messages events that occur in $h+1$ between $s(h)$ and $r(h+1)$ are $send(h+1, 2h), send(h+1, 2(h+1))$, which implies $j \in {h, h+1}$. The event $send(h+1, 2(h+1))$ does not have an incoming edge in the calculations-dependency graph. The predecessors of $send(h+1, 2h)$ in that graph are $s(h) \to_c recv(h+1, 2h) \to_c send(h+1, 2h)$. In particular, there is no path in the graph from $s(h')$ to any of these events, in contradiction to $s(h') \leadsto_c send(h+1, 2j) \leadsto_c r(h')$.
\end{proof}

\begin{lemma}
    \label{Sqrt:SequentialLockGeneralized}
    Let $h,h+i$ be two honest processors on the same segment, and let $h'$ be
    another honest processor not between $h$ and $h+i$.
    If $s(h) \leadsto s(h') \leadsto_c r(h') \leadsto r(h+i)$
    then $h$ is unvalidated or $h+i$ is unvalidated (possibly both).
\end{lemma}
\begin{proof}
    From Lemma \ref{Sqrt:StoRcalc}, there exists a validation message event
    $send(h+i, 2j)$ such that $s(h') \leadsto_c send(h+i, 2j) \leadsto_c
    r(h')$. Assume by contradiction that both $h,h+i$ are validated. The only
    validation messages events that occur in $h+i$ between $s(h)$ and $r(h+1)$
    are sending the validation values of the processors $h,h+1,..h+i$,
    formally $send(h+1, 2h), send(h+1, 2(h+1)), ..., send(h+1, 2(h+i))$.
    The predecessors of these events in the calculations-dependency graph are
    only
    $s(h+j, 2(h+j'))$ for every $0 \leq j \leq j' \leq i$ (and some receive
    events).
    In particular,
    there is no path in $Gc^D_x$ from $s(h')$ to any of these events, in
    contradiction to $s(h') \leadsto_c send(h+i, 2j) \leadsto_c r(h')$.
\end{proof}

\begin{lemma}
    \label{Sqrt:SingleLock}
    Let $h, h'$ be validated honest processors. If $s(h) \leadsto s(h')$ then $r(h) \leadsto r(h')$
\end{lemma}
\begin{proof}
    By Lemma \ref{Sqrt:StoR} there exists an event $send(h, j)$ such that $s(h') \leadsto send(h, j) \leadsto r(h')$. Then we get $s(h) \leadsto s(h') \leadsto send(h, j)$. Therefore, by Lemma \ref{Sqrt:SimpleLock}, $r(h) \leadsto send(h,j)$. So we are done by transitivity $r(h) \leadsto send(h,j) \leadsto r(h')$.
\end{proof}

\begin{lemma}
    \label{Sqrt:OneOfThree}
    Let $h,h'$ be two honest processors, if $h' \ne h+1$ and $s(h) \leadsto s(h') \leadsto s(h+1)$, then at least one of $h,h+1,h'$ is unvalidated.
\end{lemma}

\begin{proof}
    Assume by contradiction that all three of them are validated. From Lemma
    \ref{Sqrt:SingleLock}, we get $r(h') \leadsto r(h+1)$, so $s(h) \leadsto
    s(h') \leadsto_c r(h') \leadsto r(h+1)$.
    Fromm Lemma \ref{Sqrt:SequentialLock} we get contradiction.
\end{proof}

\begin{lemma}
    \label{Sqrt:OneOfThreeGeneralized}
    Let $h,+i'$ be two honest processors on the same honest segment. Let $h'
    $ be another honest processor, not between $h$ and $h+i$. If $s(h) \leadsto
    s(h') \leadsto s(h+i)$, then at least one of $h,h+i,h'$ is unvalidated.
\end{lemma}

\begin{proof}
    Assume by contradiction that all three of them are validated. From Lemma
    \ref{Sqrt:SingleLock}, we get $r(h') \leadsto r(h+i)$, so $s(h) \leadsto
    s(h') \leadsto_c r(h') \leadsto r(h+i)$.
    Fromm Lemma \ref{Sqrt:SequentialLockGeneralized} we get contradiction.
\end{proof}

\newcommand{\partA}{U^{D,1} \cup \bigcup_h q_h(U_h^{D,1})}
\newcommand{\partB}{Async^D \backslash (\partA)}

Given a deviation $D$, and an honest processor $h$, we want to ignore inputs
for which the deviation ``guesses'' some validation value - i.e.,
the data of $r(h)$ is not calculated based on the data
of $s(h)$, formally - $r(h) \not\leadsto_{c,x} s(h)$. We divide such inputs
into three types and refer directly to two of the three types. The first type
of
such inputs is defined below.
\[ U_h^{D,1} := \{x\in Async^D ~|~ s(h) \not\leadsto_{c,x} r(h), ~ r(h),
s(h)\in F^D_x, data_x(r(h))=data_x(s(h)) \}\]
Define $U^{D,1} := \bigcup_{h \in Honest} U_h^{D,1} $.\\
For notation simplicity, denote $q_h(A) := \bigcup_{a\in A}q_h(a)$\\
Next, define
\[\chi_1^D := Async^D \cup \bigcup_h q_h(U_{h}^{D, 1}) \]
Recall a deviation is defined by $2n$ functions for each adversary. The behavior of the adversary $a_i$ is defined by the functions $(func^D_i(j))_{j=1}^{2n}$ where $func^D_i(j)$ receives a list of $j$ incoming messages (a history) and outputs a list (possibly empty) of outgoing messages.\\

Next, we partition the deviations into equivalence classes according
to their behavior over short histories.
\begin{definition}
    Two deviations $D, D'$ are \textbf{equivalent}, $D \approx D'$ if for all adversary $a_i \in C$ and for all $j \leq M_0$: $func^D_i(j) = func^{D'}_i(j)$.
\end{definition}
The equivalence defined above is a proper equivalence relation.
Denote the equivalence class of a deviation $D$ with $[D]$. Denote
the set of all equivalence classes with $ Y $.

\begin{lemma}
    \label{Sqrt:FDXEquiv}
    Let $D, D'$ be deviations, if they are equivalent $D \approx D'$ then for
    every $x\in \chi$: $F^D_x = F^{D'}_x$.
\end{lemma}

\begin{proof}
    Let $x \in \chi$. Recall $F^D_x = \{e \in Events ~|~ \forall i \textup{
    s.t. } l_i\geq 1: send(a_i, M_0) \not\leadsto_x e \}$. Observe the
    calculations-dependency graph induced to $F^D_x$. It does not contain any
    event of the form $send(a_i, M_0)$, therefore every adversary has at most
    $M_0$ incoming messages in each node on that graph, i.e., that graph
    contains at most $M_0$ $ recv $ events for each adversary. Therefore, it is
    determined only by $(func^D_i(j))_{j=1}^{M_0}$.\\
    In particular, we have $F^D_x = F^{D'}_x$.
\end{proof}

\begin{lemma}
    \label{Sqrt:Chi1Equiv}
    Let $D, D'$ be deviations, if they are equivalent $D \approx D'$ then $\chi_1^D = \chi_1^{D'}$.
\end{lemma}

\begin{proof}
    By Lemma \ref{Sqrt:FDXEquiv}, $F^D_x=F^{D'}_x$, therefore $Async^D =
    Async^{D'}$.\\
    More, notice that the definition of $U_h^{D,1}$ depends only on $F^D_x$
    and the calculations-dependency graph induced on it $Gc^D_x|_{F^D_x}$ -
    therefore we
    have $U_h^{D,1}=U_h^{D',1}$ for every $h$ and for every $x$. Therefore
    $\chi_1^D = \chi_1^{D'}$.
\end{proof}

Now, similarly to the definition of $U_h^{D,1}$ we define the second
type of inputs with unvalidated processors.
\[ U_h^{D,2} := \{x\in Async^D ~|~ s(h) \not\leadsto_{c,x} r(h); s(h)\notin
F^D_x; data_x(r(h))=data_x(s(h)) \}\]
\[U^{D,2} := \bigcup_{h \in Honest} U_h^{D,2}\]

\begin{lemma}
    \label{Sqrt:qOfD2inAsync}
    For every $h$, \[q_h(U_h^{D,2}) \subseteq \chi_1^D \]
\end{lemma}
\begin{proof}
    Let $x' \in q_h(U_h^{D,2})$, so there exists $x\in U_h^{D,2}$ such that $x'
    \in q_h(x)$. Therefore, $s(h)\notin F^D_x$ therefore, $F^D_x=F^D_{x'}$.\\
    Since $x\in Async^D$, we have $nr \notin F^D_x=F^D_{x'}$ which means $x'
    \in Async^D \subseteq \chi_1^D$.
\end{proof}

Observe the definitions of $U_h^{D,1}$ and $U_h^{D,2}$, notice that
we address only two cases that do not cover all the options for which
$h$ is unvalidated. We address the case $r(h),s(h) \in F^D_x$ and
the case $s(h)\notin F^D_x$. So the case $s(h)\in F^D_x, r(h) \notin
F^D_x$ is missing. However, it turns out that it is enough when
considering inputs that break synchronization strongly as inputs in
$Async^D$. The following lemma shows that $U^{D,2}$ and $U^{D,1}$
indeed cover all the non-failing inputs in $Async^D$.

\begin{lemma}
    \label{Sqrt:AllUnvalidated}
    $Async^D \cap NoFail \subseteq U^{D,1} \cup U^{D,2}$
\end{lemma}
\begin{proof}
    Let $x \in Async^D \cap NoFail^D$. First, we show that there are at least $k+1$ unvalidated processors. Assume by contradiction that there exist $n-2k$ validated honest processors $Val = \{h_1, ..., h_{n-2k}\}$. From Lemma \ref{Sqrt:Simultaneous}, the events $s(h_1), ..., s(h_{n-2k})$ are well-ordered by the happens-before graph.   W.l.o.g assume that $s(h_1) \leadsto ... \leadsto s(h_{n-2k})$.\\
    According to Lemma \ref{Sqrt:SequentialSegment}(3), we get that for every honest segment $I=\{b, b+1, b+2, ..., c\}$: $s(b) \leadsto s(b+1) \leadsto s(b+2) .... \leadsto s(c)$. And also by \ref{Sqrt:SequentialSegment}(2) $r(b) \leadsto r(b+1) \leadsto r(b+2) .... \leadsto r(c)$. Now, let $I\cap Val=\{b_1, .. b_t\}, b_1 < b_2 ... < b_t $ be the validated processors in the segment $I$, so using the above with Lemma \ref{Sqrt:SequentialSegment}(1) we obtain $s(b_1) \leadsto r(b_1) \leadsto s(b_2) \leadsto r(b_2) \leadsto ...  \leadsto r(b_t)$.\\
    From Lemma \ref{Sqrt:OneOfThreeGeneralized}, we get that the honest
    segments are continuous in $Val$, that is for each honest segment $I$ we
    get $I \cap Val =\{h_j, h_{j+1}, h_{j+2}, ... \}$ for some $j$. Observe
    some transition from segment $I1$ to segment $I2$ - that is, $h_j \in I1,
    h_{j+1} \in I2$. From Lemma \ref{Sqrt:SingleLock}, we get $s(h_j) \leadsto
    r(h_{j+1})$.\\
    Therefore, for each honest segment that contains $t$ validating processors, there are $t-1$ disjoint paths in the calculations-dependency graph (from some $s(h_j)$ to $r(h_j)$ or to $r(h_{j+1})$ in case of transition). More, these paths are disjoint when considering the paths from all segments. For every adversary $a_i$, by Lemma \ref{Sqrt:StoRcalc}, each such path contains an event of the form $send(a_i, 2t)$.\\
    Observe the event $nr=send(a_1, 2(n-l))$. Let $h_j\in Val$ be the latest validating processor such that $s(h_j) \leadsto nr$. Observe the disjoint paths described above that happen before $s(h_j)$. There are at least $j-k$ such paths. However there are at most $n-l$ messages of the form $send(a_1, 2t)$ before $s(h_j)$ in the happens before graph. Therefore, $j-k\leq n-l \Longrightarrow j \leq n - l + k (*)$. \\
    On the other hand, since $x\in Async^D$, we have $send(a_i, M_0) \leadsto nr$ for some adversary $a_i$. From Lemma \ref{Sqrt:StoRcalc} there is some event $s(h_{j+1}) \leadsto send(a_1, 2t) \leadsto r(h_{j+1})$. From maximality of $h_j$ and total order over the events of $a_1$, we get $nr \leadsto send(a_1, 2t) \leadsto r(h_{j+1})$. So there are at least $n-2k - (j+1) -k$ disjoint paths as described above that happen after $nr$. And therefore also after $send(a_i, M_0)$. Therefore there are at most $n-\frac{M_0}{2} $ events of the form $send(a_i, 2r)$ that happen after $send(a_i, M_0)$. So $n-2k-(j+1)-k \leq n - M_0$. Recall $\frac{M_0}{2}= n -l + 4k + 2, $ so $n - l + k + 1 \leq j$. In contradiction to $ (*) $.\\
    So there are at least $k+1$ unvalidated processors in $x$. But there are only $k$ honest segments, therefore there exist two unvalidated processors $h, h'$ on the same honest segment.\\
    Since $x\in NoFail^D$, we have $data(s(h))=data(r(h)),
    data(s(h'))=data(r(h'))$. Assume by contradiction that $x \notin U^{D,1}
    \cup U^{D, 2}$. So $s(h), s(h') \in F^D_x, r(h), r(h') \notin F^D_x$.
    W.l.o.g assume $h$ is located before $h'$ on their segment. So by Lemma
    \ref{Sqrt:SequentialSegment} we get $r(h) \leadsto s(h+1) \leadsto s(h+2)
    \leadsto ... \leadsto s(h') \in F^D$. So $r(h) \in F^D$ contradiction.
\end{proof}
\begin{lemma}
    \label{Sqrt:IgnoreChi1}
    \[Pr(x \in NoFail^D ~|~ x\in \chi_1^D) \leq \frac{1}{n}\]
\end{lemma}
\begin{proof}
    \[ Pr(NoFail^D ~|~ \chi_1^D) = \frac{| NoFail^D \cap \chi_1^D|}{| \chi_1^D
    |} \leq\]
    By Lemma \ref{Sqrt:AllUnvalidated},
    \[\leq\frac{| U^{D,1} \cup U^{D,2} |}{| \chi_1^D|} \leq \]
    Union bound the nominator,
    \[\leq \frac{| U^{D,1}|}{|\chi_1^D|} +  \frac{|U^{D,2}|}{| \chi_1^D|}
    \leq \]
    By definition of $\chi_1^D$ and by Lemma \ref{Sqrt:qOfD2inAsync},
    \[\leq \frac{| U^{D,1}|}{|U^{D,1}\cup \bigcup_h{q_h(U_h^{D,1})}|} +  \frac{|
    U^{D,2}|}{|U^{D,2}\cup \bigcup_h{q_h(U_h^{D,2})}|} \leq \]
    Let $h1 := argmax_h |U_h^{D,1}|$. So, $|U^{D,1}| \leq n
    |U_{h1}^{D,1}|$
    and $|U^{D,1}\cup \bigcup_h{q_h(U_h^{D,1})}| \geq m |U_{h1}^{D,1}|$.
    Similarly, define $h2 := argmax_h |U_h^{D,2}|$ and deduce the analog
    inequalities. This gives us the bound,
    \[\leq \frac{n |U_{h1}^{D,1}|}{m |U_{h1}^{D,1}|} + \frac{n
    |U_{h2}^{D,2}|}{m |U_{h2}^{D,2}|} = \frac{2n}{m} = \frac{1}{n} \]
\end{proof}

Next, we define the operation of $a_1$ until it hits the point of no
return as a function.

\begin{definition}
    Let $D$ be a deviation, define the operation of $a_1$ as the function $g_D: \chi\backslash \chi_1^D \to [n]^{n}\times [m]^{n-l} $, where $g_D(x)$ is the series of the first $2(n-l)$ outgoing messages of $a_1$, concatenated with the data values of the $l$ honest processors that follow $a_1$ on the ring $\{a_1+1,... a_1+l\} \subseteq I_1$.
\end{definition}
The function $g_D$ is well-defined because for every $x\in \chi\backslash \chi_1^D$ $x\notin \chi_1^D \Longrightarrow x\notin Async^D \Longrightarrow nr \in F^D_x$ so the event $nr$ happens in the execution of $x$.\\

Note that the first $2(n-l)$ outgoing messages of $a_1$ are perceived by honest processors in $I_1$ as $\{d_i\}_{i=a_1-n+l+1 \pmod n}^{a_1}, \{v_i\}_{i=1}^{n-l}$. Recall that for every $x\in NoFail^D$, the last $l$ outgoing data messages of $a_1$ are the data values of the processors $\{a_1+1,.. a_1+l\}$ since $l_1 \geq l$. Overall, we get $\DataValues, \{v_i\}_{i=1}^{n-l}$. So for every $x\in NoFail^D$, the output calculated by honest processors in $I_1$ is $f(g_D(x))$.\\

For all the inputs in $g_D^{-1}(g_D(x)) \cap NoFail^D$, the outgoing
messages of $a_1$ are $g_D(x)$ and some extra $l$ validation
messages. Since distinct inputs $x, x'\in NoFail^D$ define different
outgoing messages for $a_1$, we get that $|g_D^{-1}(g_D(x)) \cap
NoFail^D | \leq n^l$. We denote the inputs for which it does not
hold as $\chi_2^D$ and show they are irrelevant as well.

\begin{definition}
    Define $out_1^D: \chi \to [n]^{n}\times [m]^n \cupdot \{\emptyset\}$. If $a_1$ sends $2n$ messages, define $out_1^D(x)$ to be the list of outgoing messages of $a_1$ . Otherwise, define $out_1^D(x)=\emptyset$.
\end{definition}

\begin{lemma}
    \label{Sqrt:OutIsInjective}
    For every $x,x'\in NoFail$:
    \[out_1^D(x)=out_1^D(x') \Rightarrow x=x'\]
\end{lemma}
\begin{proof}
    Let $D$ be a deviation. Let $x, x'\in NoFail^D$. Since $x, x' \notin NoFail^D$, $out_1^D(x) \ne \emptyset , out_1^D(x') \ne \emptyset$. Assume $out_1(x)=out_1(x')$, then the last $l_1=|I_1|$ outgoing messages of $a_1$ are equal in the execution of $x$ and in the execution of $x'$. Since $x,x'\in NoFail^D$ then the data value of $I_1$ are identical in $x$ and $x'$. Similarly, the validation values of $I_1$ are identical in $x$ and $x'$. Since the outgoing messages of (the last processor in) $I_1$ are determined by $I_1$'s data values, validation values and outgoing messages of $a_1$, the outgoing messages of $I_1$ are equal in $x$ and $x'$. However, these are the incoming messages of $a_2$, so $out_2^D(x)=out_2^D(x')$.\\
    Applying the same argument inductively, one can induce that $\forall j: out_j^D(x)=out_j^D(x')$ and that the validation and data values of $I_j$ are identical in $x$ and $x'$ for all $j$. Which means $x=x'$.
\end{proof}

\begin{definition}
    \[\chi_2^D := \{x\in \chi\backslash \chi_1^D ~|~ m^{l+1} < |g_D^{-1}(g_D(x))| \}\]
    \[\chi_3^D := \chi\backslash (\chi_1^D \cupdot \chi_2^D)\]
\end{definition}

Next, we show that inputs in $\chi_2^D$ are irrelevant as well.
\begin{restatable}{lemma}{IgnoreChi2}
    \label{Sqrt:IgnoreChi2}
    $Pr(x\in NoFail^D ~|~ x\in \chi_2^D) \leq \frac{1}{n}$
\end{restatable}
\begin{proof}
    Let $y\in g_D(\chi_2^D)$. For every $x\in g_D^{-1}(y) \cap NoFail^D$, we have $out_1^D(x) = <y, s>$ for some $s \in [m]^l$ where $<*, *>$ denotes concatenation. Therefore, $|\{out_1^D(x) ~|~ g_D(x) = y, x \in NoFail^D \}| \leq m^l$. But $out_1^D$ is injective on non-failing inputs by Lemma \ref{Sqrt:OutIsInjective} therefore $|g_D^{-1}(y) \cap NoFail^D| \leq m^l$. But $y\in g_D(\chi_2^D)$, so $|g_D^{-1}(y)| \geq m^{l+1}$, therefore $Pr(NoFail^D ~|~ g_D^{-1}(y)) \leq \frac{1}{n}$.\\
    Overall,
    \[Pr(NoFail^D ~|~  \chi_2^D) = \sum_{y\in g_D(\chi_2^D)} Pr(NoFail^D ~|~ g_D^{-1}(y)) Pr(g_D^{-1}(y)) \leq \frac{1}{n}\]
\end{proof}

\begin{lemma}
    \label{Sqrt:ChisAndGEquiv}
    If $D\approx D'$, then $\equiv g_{D'}$, $\chi_2^D=\chi_2^{D'}, \chi_3^D=\chi_3^{D'}$.
\end{lemma}
\begin{proof}
    Assume $D\approx D'$, then by Lemma \ref{Sqrt:Chi1Equiv}, we get $\chi_1^D = \chi_1^{D'}$. Therefore from the definitions of $\chi_2^D$ and $\chi_3^D$, it is enough to show that $g_D \equiv g_{D'}$.\\
    For every $x\in \chi \backslash \chi_1^{D}$, we have $x \notin Async^D$, therefore $nr \in F^D_x$. So the whole calculation tree of $nr$ is determined by the class $[D]$ and is identical in $D$ and $D'$. Therefore, the first $2(n-l)$ outgoing messages of $a_1$ are also identical. Therefore $g_D(x)=g_{D'}(x)$.
\end{proof}

So given an equivalence class $[D]$, one may refer to $g_D, \chi_1^D, \chi_2^D, \chi_3^D$ w.l.o.g.\\
Now, we calculate a naive bound over the amount of equivalence
classes.
\begin{lemma}
    \label{Sqrt:BoundEqClass}
    $|Y| \leq exp(n^{n - l + 4k + 6}m^{n - l + 4k + 2})$
\end{lemma}

\begin{proof}
    Let us count the number of possible $func_i(j)$. The domain of $ func_i(j) $ is of size $n^{j/2} m^{j/2}$ as each input consists of $\frac{1}{2}j$ data messages and $\frac{1}{2}j$ validation messages. The range of $func_i(j)$ is of size $\sum_{len=1}^{n} n^{len} m^{len} < m^{2n+1} = exp(\log(m)(2n+1)) < exp(n^2)$ all possible series of outgoing messages. Therefore, the number of options for $func_i(j)$ is at most $exp(n^2)^{n^{j/2}m^{j/2}}=exp(n^{j/2+2}m^{j/2})$.\\
    An equivalence class is defined by $func_i(j)$ for $k$ adversaries, $j \leq M_0$. So \[|Y| \leq exp(n^{M_0/2+2}m^{M_0/2})^{k M_0} \leq exp(n^{M_0/2+2}m^{M_0/2})^{n^2} = exp(n^{M_0/2+4}m^{M_0/2})=\]
    \[ = exp(n^{n-l+4k+2+4}m^{n-l+4k+2}) = exp(n^{n-l+4k+6}m^{n-l+4k+2})\]
\end{proof}
Note that all of the definitions and claims above do not rely on $f$, so they hold for every selection of $f$.\\
Next, we introduce a bias property of a deviation with respect to a
function $f$.

\begin{definition}
    A deviation $D$ is \textbf{$\epsilon$-good} with respect to a function $f$, if $Pr(outcome=1) > \frac{1}{n}+\epsilon$
\end{definition}

We want to show that with high probability over selecting $f$, for
every deviation the bias property above does not hold. We prove it
by first inducing a similar bias property on the equivalence class
of $D$, second bounding the probability that an equivalence class
upholds the bias property for a random function $f$, and finally
applying a union bound over all the equivalence classes.

\begin{definition}
    An equivalence class $[D]$ is \textbf{$\epsilon$-good} with respect to a function $f$, if $Pr(f(g_D(x))~|~ x\in \chi_3^D) > \frac{1}{n}+\epsilon$
\end{definition}
\begin{lemma}
    \label{Sqrt:ReduceToEquivalence}
    If an adversarial deviation $D$ is $\epsilon$-good with respect to $f$ then $[D]$ is $\epsilon$-good with respect to $f$ and also $\frac{|\chi_3^D|}{|\chi|}> \epsilon$.
\end{lemma}
\begin{proof}
    Let $\epsilon > 0$.\\
    By the law of total probability, for every deviation $D$ and every function $f$:
    \[Pr(outcome=1) = \sum_{i=1}^3Pr(outcome=1 ~|~ x\in \chi_i^D) Pr(x \in \chi_i^D)\]
    By Lemmas \ref{Sqrt:IgnoreChi1}, \ref{Sqrt:IgnoreChi2}, we know that for $i=1,2$: $Pr(NoFail^D ~|~ x\in \chi_i^D) \leq \frac{1}{n}$. And since $\{x \in \chi ~|~ outcome=1\} \subseteq NoFail^D$, we get
    \[Pr(outcome=1) \leq \frac{1}{n}Pr(x\notin \chi_3^D) + Pr(outcome=1 ~|~ x\in \chi_3^D)Pr(x \in \chi_3^D) \]
    So if $Pr(outcome=1) > \frac{1}{n} + \epsilon$ then $Pr(outcome=1 ~|~ x\in \chi_3^D)  > \frac{1}{n} + \epsilon$ and also $Pr(x\in \chi_3^D) > \epsilon$.
    Since, $Pr(x\in \chi_3^D) > \epsilon$, we get $\frac{|\chi_3^D|}{|\chi|} > \epsilon$.\\
    If $outcome=1$, then every processor calculates $output=1$, in particular the processors in $I_1$ calculate $output = f(g_D(x)) = 1$. Therefore
    \[ Pr(f(g_D(x) ~|~ x\in \chi_3^D) \geq Pr(outcome=1 ~|~ x\in \chi_3^D)  > \frac{1}{n} + \epsilon \]
    I.e., the class $[D]$ is $\epsilon$-good with respect to $f$.
\end{proof}

The next Lemma is due to Chernoff's concentration bound of the
Binomial distribution.
\begin{lemma}
    \label{Sqrt:Conentration}
    Let $\{c_i\}_{i\in J} \subseteq \mathbb{N}$ be a bounded series of natural numbers, $c_i \leq C$. Denote $\sum{c_i} = S$. Let $0<p<1,~\epsilon > 0$. Let $\{b_i(p)\}_{i \in J}$ be a series of i.i.d Bernoulli random variables. Then,
    \[
    Pr(\frac{\sum_{i\in J}{c_i\cdot b_i}}{S} > p+\epsilon) \leq exp(-2 \epsilon^2 \frac{S}{C})
    \]
\end{lemma}
\begin{proof}
    We reformulate and apply Hoeffding's inequality
    \[Pr(\frac{\sum_{i\in J}{c_i\cdot b_i}}{S} > p+\epsilon) = Pr(\frac{\sum_{i\in J} c_i b_i}{|J|} - \frac{Sp}{|J|} > \frac{S\epsilon}{|J|} ) =\]
    Denote $X_i=c_i b_i$, $\mean{X_i} = \frac{\sum_{i\in {J}} X_i}{|J|}$, $t= \frac{S\epsilon}{|J|}$. Substitute
    \[= Pr(\mean{X_i} - E[\mean{X_i}] > t) \]
    Apply Hoeffdding's inequality
    \[\leq exp(-\frac{2 |J|^2 t^2}{\sum_i (c_i - 0)^2}) = exp(-\frac{2S^2\epsilon^2}{\sum_i c_i^2}) (*) \]
    The bound above achieves maximum when $\sum_i c_i^2$ achieves maximum
    \begin{equation*}
    \begin{aligned}
    & \underset{\{c_i\}_{i\in J}}{\text{maximize}}
    & & \sum_i c_i^2 \\
    & \text{subject to}
    & & 0 \leq c_i \leq C, \; i \in J
    & & \sum_i c_i = S
    \end{aligned}
    \end{equation*}
    The maximum is attained at $c_i = C$ for $\frac{S}{C}$ indices, and $c_i=0$ for the rest (assuming $|J| > \frac{S}{C}$).
    So the maximum value is bounded with $\frac{S}{C}C^2=S C$ which means
    \[(*) \leq exp(-\frac{2S^2\epsilon^2}{S C}) = exp(-2\frac{S}{C}\epsilon^2) \]

\end{proof}

\begin{lemma}\label{Sqrt:BoundBinomial}
    For every equivalence class $[D]$, if $\frac{|\chi_3^D|}{|\chi|} > \epsilon $ then
    \[
    Pr_f([D]\textup{ is }\epsilon\textup{-good}) \leq exp(-2 \epsilon^3 \cdot N)
    \]
    for $N= n^{n-k}m^{n-k-l-1}$
\end{lemma}
\begin{proof}
    Let $D$ be a deviation such that $\frac{|\chi_3^D|}{|\chi|} > \epsilon$. Let $f$ be a function that defines the protocol such that the class $[D]$ is $\epsilon$-good (with respect to $f$).\\
    So $Pr(f(g_D(x)) ~|~ x\in \chi_3^D) > \frac{1}{n} + \epsilon$. We reformulate the inequality and apply Lemma \ref{Sqrt:Conentration}
    \[Pr(f(g_D(x)) ~|~ x\in \chi_3^D) = \frac{\sum_{x\in \chi_3^D, f(g_D(x)=1)}1}{|\chi_3^D|} = \frac{\sum_{y; f(y)=1}\sum_{x\in \chi_3^D \cap g_D^{-1}(y)}1}{|\chi_3^D|} =\]
    \[ = \frac{\sum_{y}|\chi_3^D \cap g_D^{-1}(y)| \mathbbm{1}[f(y)=1]}{|\chi_3^D|}\]
    Denote $c_y = |\chi_3^D \cap g_D^{-1}(y)|, S := \sum_y |\chi_3^D \cap g_D^{-1}(y)| = |\chi_3^D| $
    \[= \frac{\sum_y c_y \mathbbm{1}[f(y)=1]}{S}\]
    Notice that from the definition of $\chi_3^D$ we get $C := m^{l+1} \geq c_y 
    $, and that $\mathbbm{1}[f(y)=1]$ is $Bernouli(\frac{1}{n})$. So we can 
    apply Lemma \ref{Sqrt:Conentration} and obtain
    \[Pr_f([D]\textup{ is } \epsilon\textup{-good}) = Pr_f(Pr(f(g_D(x)) ~|~ x\in \chi_3^D) > \frac{1}{n} + \epsilon) = Pr_f(\frac{\sum_y c_y \mathbbm{1}[f(y)=1]}{S} > \frac{1}{n} + \epsilon ) \leq \]
    \[ \leq exp(-2 \epsilon^2 \frac{S}{C}) = exp(-2 \epsilon^2 
    \frac{|\chi_3^D|}{m^{l+1}} ) \leq exp(-2 \epsilon^2 \epsilon |\chi| 
    m^{-l-1}) = \]
    \[exp(-2 \epsilon^3 n^{n-k}m^{n-k-l-1}) \]
\end{proof}

\SquareResilienceTheorem*
\begin{proof}
    It is enough to show resilience for the maximal $k$ and the minimal 
    $\epsilon$, so it is enough to prove the protocol is \kunbiased for 
    $k=\frac{1}{10}\sqrt{n}, \epsilon = n^{-\sqrt{n}}$.\\
    Let $D$ be a deviation. If $Pr(\chi_3^D) < \epsilon$ then by Lemma \ref{Sqrt:ReduceToEquivalence}, it is not $\epsilon$-good and we are done for every $f$.
    Otherwise, $Pr(\chi_3^D) > \epsilon$.
    So when we select a random function $f$,
    \[Pr_f(\exists \textup{ deviation } D \textup{ such that }Pr(outcome=1) > \frac{1}{n}+\epsilon) = \]
    Then by Lemma \ref{Sqrt:ReduceToEquivalence}, we have
    \[= Pr_f(\exists D \textup{ that is } \epsilon\textup{-good}  \wedge \frac{|\chi_3^D|}{|\chi|} > \epsilon) \]
    From Lemma \ref{Sqrt:ReduceToEquivalence} it is enough to consider the equivalence classes
    \[
    Pr_f(\exists D \textup{ that is } \epsilon\textup{-good}\wedge \frac{|\chi_3^D|}{|\chi|} > \epsilon) \leq
    Pr_f(\bigcup_{[D]\in Y;\frac{|\chi_3^D|}{|\chi|} >\epsilon}\{[D]\textup{ is }\epsilon\textup{-good}\}) \leq\]
    Using a union bound over $Y$,
    \[
    \leq \sum_{[D]\in Y; \frac{|\chi_3^D|}{|\chi|} > \epsilon} Pr_f([D]\textup{ is }\epsilon\textup{-good})
    \]
    By Lemma \ref{Sqrt:BoundBinomial} $Pr_f([D]\textup{ is }\epsilon\textup{-good})$ is small, and by Lemma \ref{Sqrt:BoundEqClass} $|Y|$ is small
    \[
    \leq exp(n^{n-l+4k+6}m^{n-l+4k+2}) exp(-2\cdot \epsilon^3 \cdot N) <
    \]
    Recall $N=n^{n-k}m^{n-k-l-1}, k = \frac{1}{10}\sqrt{n}, \epsilon = 
    n^{-\sqrt{n}}=n^{-10k}$, $l=10\sqrt{n}=100k, m = 2n^2<n^3$ \\
	So we get by straight calculation,
    \[
    < exp(n^{n-l}m^{n-l}(n^{4k+6}m^{4k+2} - n^{-31k}n^{l-k}m^{-k-1}))<
    exp(n^{n-l}m^{n-l}(n^{16k} - n^{-31k+100k-k-6k})) =\]
    Which is very small as required.
\end{proof}

\textbf{Remark:} Asymptotically, our resilience result for
\PhaseAsyncLead is tight because with high probability over
selecting $f$ the protocol is not \kresilient for $k=\sqrt{n}+3$. A
straight-forward variation of the naive rushing attack works for
\PhaseAsyncLead as well, namely, When handling validation messages,
the adversaries send them as defined in the protocol.

During the first $n-k$ rounds, when handling data messages, the
adversaries behave like pipes. Thus, within $n-k$ rounds each
adversary knows the data values of all the honest processors and
also the first $n-l$ validation values. Therefore, if the honest
segments are of length $< k-3$, then each adversary can control 3
entries in the input of $f$. Thus, for a random $f$, the adversaries
can control $f$ almost for every input.

It still remains to determine whether a similar result holds for
every $f$, which is a stronger notion than ``with high probability''
over selecting $f$.

\textbf{Remark:} The protocol above is defined only given a function
$f$, so its definition is huge. This can be solved by rephrasing the
protocol. Instead of including $f$ as part of the protocol's
definition, each processor iterates (locally) over all the possible
functions $f$ and selects the first for which every deviation is not
$\epsilon$-good.

\textbf{Remark:} The protocol defined above is not a fair leader
election protocol because even without any deviation we have only
$Pr(outcome=j)\approx \frac{1}{n}$ and not strict equality as
required by the definition. However, a significant fraction of the
functions gives $\forall j: Pr(outcome=j)=\frac{1}{n} $. We did not verify that 
the proof can be augmented to consider only such functions.

\newpage
\subsomething{PhaseAsyncLead Pseudo-Code}
\label{Sqrt:Pseudo-Code}
For notation simplicity, we specify different handler functions for validation messages and for data messages. In practice, each processor treats all the odd incoming messages (first incoming message, third incoming message etc.) as data messages and all the even incoming messages (second, fourth etc.) as validation messages. We assume implicitly that if a processor receives a validation message as an odd message, then it aborts. Similarly, if it receives a data message as an even message, then it aborts.

For readability, we write ``Send $DataMessage(d)$'', in practice it could be just ``Send $d$''. Same goes for ``Send $ValidationMessage(v)$''.

\begin{algorithm}[H]
    \caption{\PhaseAsyncLead. Synchronized Resilient Leader Election On Ring. Code for a $normal$ processor $i$.}
    \Fn{Init()}{
        Init arrays $d[1...n]$, $v[1,...n]$\;
        $d[i] = Uniform([n])$\;
        $buffer = d[i]$\;
        $round=0$\;
    }
    \Fn{UponRecieveDataMessage($dataValue$) processor $i$}{
        Send $DataMessage(buffer)$\;
        $round\plusplus$\;
        \tcp{Delay the incoming message until the next round}
        $buffer = dataValue$\;
        $j=i-round\pmod n$\;
        $d[j] = dataValue$\;
        \If{$round==i$}{
            $v[i] = Uniform([n])$\tcp*{Randomize the validation value}
            Send $ValidationMessage(v[i])$
        }
        \If{$round==n$}{
            \lIf{$value \ne d[i]$}{$Terminate(\bot)$}
        }
    }
    \Fn{UponRecieveValidationMessage($validationValue$)}{
        \uIf{$round==i$}{
            \tcp{Validate the validation message}
            \lIf{$v[i] \ne validationValue$}{$Terminate(\bot)$}
        }\Else{
        $v[round] = validationValue$\;
        Send $ValidationMessage(validationValue)$\;
    }
    \If{$round==n$}{
        $output=f(d[1], ..., d[n], v[1], ..., v[n-l])$\;
        $Terminate(output)$
    }
}
\end{algorithm}
\newpage
\begin{algorithm}[H]
    \caption{\PhaseAsyncLead. Synchronized Resilient Leader Election On Ring. Code for $origin$ processor. $i=1$}
    \Fn{Init()}{
        Init arrays $d[1...n]$, $v[1,...n]$\;
        $d[1] = Uniform([n])$\tcp*{Data value}
        Send $DataMessage(d[1])$\;
        $round=1$\;
        $v[1] = Uniform([n])$\tcp*{Validation value}
        Send $ValidationMessage(v[1])$\;
        $buffer = \bot$ \tcp*{Global variable}
    }
    \Fn{UponRecieveDataMessage($value$)}{
        $buffer=value$\;
    }

    \Fn{UponRecieveValidationMessage($value$)}{
        \uIf{$round==1$}{
            \lIf{$v[1] \ne validationValue$}{$Terminate(\bot)$}
        }
        \Else{
            Send $ValidationMessage(value)$
        }
        Send $DataMessage(buffer)$\;
        $round\plusplus$\;
        \If{$round==n$}{
            $output=f(d[1], ..., d[n], v[1], ..., v[n-l])$\;
            $Terminate(output)$
        }
    }
\end{algorithm}

\subsomething{Motivating the need for a random function}
\label{Sqrt:RandomFunctionIsRequired}
Consider adding the phase validation mechanism to \RingOriginalProtocol
while keeping the sum function to select the final output (and not a random function).
While the phase validation mechanism keeps all the processors
synchronized, adding it to \RingOriginalProtocolSpaced makes it
non-resilient to $k=4$ adversaries. The adversaries can abuse the
validation messages to share partial sums of $S=\sum_{h \notin C}d_h
\pmod n$ quickly and thus cheat. For example, assume all honest
segments are of length $L = \frac{n-k}{4}$. As in previous attacks, the adversaries
do not select data values for themselves and they rush the honest
data values. For validation messages, adversaries behave honestly
when the round's validator is honest. So after $L$ rounds each
adversary knows the data values of the segment behind it. In
particular, each adversary $a_i$ knows the sum of these data values,
that is $a_i$ knows $S_i:=\sum_{h\in I_{i-1}} d_h \pmod n$. Moreover,
only one adversary was a round's validator, w.l.o.g it was $a_1$.
Then, when the adversary $a_2$ is the round's validator, the
adversaries can use the validation messages wisely to calculate the
total sum $S$: $a_2$ sends $S_2$, then $a_3$ sends $S_2+S_3$, then
$a_4$ sends $S_2+S_3+S_4$ and finally $a_1$ sends
$S_2+S_3+S4+S_1=S$. So now both $a_1$ and $a_2$ know the sum $S$.
Next, when $a_3$ is the round's validator, the adversaries can share
$S$: notice that when an adversary is the round's validator, any
adversary may initiate the validation process so the following deviation is
undetectable - $a_2$ sends $S$, then $a_3$ sends $S$, then $a_4$ sends $S$
and finally $a_1$ sends $S$. So all the adversaries know $S$ after
less than $3L$ rounds - $L$ rounds to learn $S_i$, then less than $L$ until $a_2$
is the validator, finally $L$ more rounds until $a_3$ is the validator. In particular, they know $S$ before committing (before sending
$n-L$ messages). Recall that each adversary $a_i$ has $k=4$ spare messages,
so just after sending $n-L-4$ messages, $a_i$ sends the value $w-S$ and $3$
zero messages. Overall, the adversaries control the sum of their outgoing
messages without getting caught. Therefore, they can control the outcome.

\mySection{Proofs for Resilience Impossibility of 
\texorpdfstring{$k$}{k}-Simulated Trees}
\label{appendix:TreesImpossibility}

In this section, we provide a proof for Theorem \ref{TreesImpossibility:LeaderElectionImpossibility}
from \SectionText \ref{section:TreesImpossibility}. W.l.o.g, we consider only
deterministic protocols by assuming every processor receives a
random string as input.

\begin{figure}
    \centering
    \includegraphics[width=0.6\textwidth]{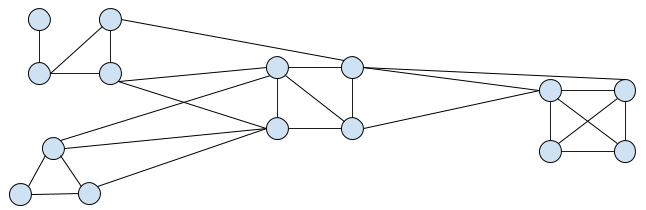}
    \caption{A $k$-simulated tree with $k=4$}
    \label{TreesImpossibility:Figure}
\end{figure}

\RestatableSimulatedTreeGraphLeaderElectionImpossibility*
\begin{proofoutline}{Theorem \ref{TreesImpossibility:LeaderElectionImpossibility}}
    First, we prove in Lemma \ref{TwoSideAssureLemma} that in our model
    there is no $\epsilon$-$1$-unbiased two-party \textit{coin toss} protocol. Then
    in Lemma \ref{TreeCoinAttack}, we conclude there is no
    $\epsilon$-$1$-unbiased coin toss protocol for every tree by induction
    on the number of vertices.
    
    Let $G$ be $k$-simulated tree, assume it is simulated by the tree
    $T$. In Lemma \ref{SimulatedGraphCoinImpossibility} we notice that
    any protocol for $G$ can be simulated by $T$ and therefore there is
    no \kunbiased coin toss protocol for $G$. Therefore by a
    variant of Lemma \ref{Model:UnbiasResilienceEquiv} for coin-toss,
    there is no $\epsilon$-$k$-resilient coin toss protocol. Finally, we
    conclude because coin toss can be reduced to fair leader election by
    taking the lower bit, as stated in \SubSomethingText
    \ref{section:ProblemsRelations}.
\end{proofoutline}

Note that for this section, we consider general protocols, i.e., not
only symmetric ones, so the strategy of a processor might depend on
its location in the graph.

First, we show that every two-party protocol cannot guarantee any
resilience.
\begin{definition}
    Let $P$ be a protocol for $n$ processors (not necessarily symmetric). We say that a coalition $ C\subset V$ \emph{assures} $o_0$ if there exists an adversarial deviation of $C$ from $P$, such that for every message scheduling, $Pr(o_0)=1$.
\end{definition}
\begin{lemma}\label{TwoSideAssureLemma}
    For a graph with two processors $V=\{A, B\}$, for every protocol $P$ with valid outputs set $\Omega=\{0, 1\}$ (i.e., a two-party coin toss protocol) that guarantees a bounded amount of messages, such that the valid inputs set is a cartesian product $I_A \times I_B$, the two following statements hold:
    \begin{enumerate}[noitemsep,topsep=0pt]
        \item Either $a$ assures $0$ or $ B $ assures $1$
        \item Either $a$ assures $1$ or $ B $ assures $0$
    \end{enumerate}
\end{lemma}
In other words, either there is a favorable value $b\in \{0, 1\}$
such that both processors assure $b$. Or, one of the processors is a
dictator which selects the output - it assures $ 0 $ and also
assures $ 1 $.

Our proof is inspired by a proof of Boppana and Narayanan \cite{boppana2000perfect}. They prove a similar result for a different model,
resilient fair coin-toss in two players is impossible in a synchronous perfect information model.

We prove the claim by induction on the maximal amount of messages in the protocol.
\begin{proof}
    By induction on the maximal number of receive events $E$ of a protocol.
    
    \par \textbf{Induction Basis}: If there are no events $E=0$, then the protocol determines the outcome without sending any message. So the outcome is either 0 or 1. If it is 1, then both $A$ and $B$ assure 1, else it is 0, so both $A$ and $B$ assure 0. Therefore, we have a favorable value.

    \par \textbf{Induction step}: There might be a receive event in the protocol, $E>0$.
    If there is an input $(i_a, i_b) \in I_a \times I_b$ for which both of the processors do not send any message at initialization, let the outcome of the protocol for such an input be $o_0$. So both of the processors assure $o_0$ by terminating at initialization. Because if one of them, say $A$ would wait to receive a message from $B$ when $B$'s input is $i_b$ then it will wait forever and therefore the protocol will fail.

    Otherwise, there is no such input. So there exists a processor that sends a message at initialization for every possible input since the inputs space is a cartesian product. Assume w.l.o.g that $A$ sends a message at initialization for every input. For every legitimate option $ M $ for $ A $'s first message, consider the protocol $P_M$ that results by $ B $ receiving $ M $ and continuing the original protocol. The inputs set for $A$ in $P_M$ is the inputs in  $I_A$ such that the first outgoing message of $A$ is $M$. The inputs set for $B$ in $P_M$ is $I_B$. And the inputs set for the $P_M$ is their cartesian product so by the induction hypothesis, in $P_M$ either $A$ assures 1 or $B$ assures 0. If there exists a legitimate value $ M_0 $ for which $ A $ assures 1 in $P_{M_0}$ then by sending $ M_0 $ at initialization, $A$ assures 1 in $P$.

    Else, for every legitimate value $ M $, $B$ assures 0 in $P_M$, so by waiting for $A$ to send its first message, $ B $ assures 0 in $P$.

    The proof second claim holds by symmetry.
\end{proof}
\textbf{Remark:} We do not build upon the impossibility proof
introduced by Abraham et al.\cite{abraham2013distributed} for two
reasons: First, our claim is stronger because we address both the ability
to assure $1$ and the ability to assure $0$. Second, their proof includes
a non-trivial reduction to a synchronous model but this reduction is
not proved.

Next, we generalize Lemma \ref{TwoSideAssureLemma} for a tree
\begin{lemma}\label{TreeCoinAttack}
    In a tree network, for every protocol that selects a value in $\{0,1\}$, such that its inputs set that is a cartesian product, there exists a processor that assures $ 1 $ or there exists a processor that assures $ 0 $.
\end{lemma}
We prove the claim by induction on the number of vertices. In the
induction step, we focus on a leaf $a$ that is connected to only to
the processor $b$. The main observation in the proof is that the
conversation between $a$ and $b$ can be viewed as a coin toss
protocol where $b$ simulates the rest of the tree.
\begin{proof}
    By induction on $n$, the number of processors in the tree.

    \par\textbf{Basis} $n=1$: A single processor assures 1.
    
    \par\textbf{Induction step}: Consider a leaf in $T=(V, E)$. Let $P=<S_x; x \in V>$ be a protocol for $T$. Let $a \in V$ be a leaf, let $b\in V$ be its neighbor. Let $T'$ be the tree after discarding $a$. Consider the following two-party protocol $P_a$ between $a$ and $b$. Let $a$ execute its strategy $S_a$, let $b$ simulate the rest of the processors. If $a$ assures some output $b\in \{0, 1\}$ in $P_a$, then it assures $b$ in $P$ and we are done. Otherwise, by Lemma \ref{TwoSideAssureLemma} $b$ is a dictator in $P_a$.

    Consider the following protocol $P'$ for $T'$. Let $b$ simulate $a$ and $b$ with $S_a$ and $S_b$ and let every other processor in $x \in V' \backslash \{b\}$ execute its strategy $S_x$. By the induction hypothesis, there exists some processor $c\in T'$ that assures a value $bit \in \{0, 1\}$ in $P'$ with a strategy $S_{bad}$. If $b \ne c$, then with the same strategy $S_{bad}$, $c$ assures $bit$ in $P$. Else, $b=c$. Now recall $b$ is a dictator in $P_a$, so by communicating towards $T'$ using $S_bad$ and communicating towards $a$ to select the output to be $bit$, the processor $b$ assures $bit$ in $P$.
\end{proof}

From Lemma \ref{TreeCoinAttack} it follows immediately that if a
graph can be $k$-simulated by a tree then a coalition of size $k$
assures some outcome.
\begin{corollary}
    \label{SimulatedGraphCoinImpossibility}
    For every graph $G=(V,E)$ that is $k$-simulated by a tree $T=(V_T,E_T)$, for every fair coin toss protocol, there exists a coalition of size $k$ that assures $1$ or assures $0$.
\end{corollary}
\begin{proof}
    Let $P$ be a fair coin toss protocol for $G$ with inputs set $\prod_{x\in V}I_x$. Let $f: V\to V_T$ be the simulation mapping. Let $P_T$ be the simulation of $P$ by $T$, where each processor $v\in V_T$ simulates $f^{-1}(v)$, and its input set is $\prod_{x\in f^{-1}(v)}I_x$. For correct simulation, annotate every message with its original source and destination in $P$. By the definition of $G$, every message from $x$ to $y$ is sent on a legitimate link in $T$, so the simulation is well-defined. Since the inputs set of $P_T$ is a cartesian product $\prod_{v\in V_T} \prod_{x\in f^{-1}(v)}I_x$, so the conditions for Lemma \ref{TreeCoinAttack} hold and there exists a processor $v_0 \in V_T$ that assures some value $bit \in \{0, 1\}$ in $P_T$.

    From the definition of $G$ and $T$, $f^{-1}(v_0)$ is connected in $G$ so the coalition $f^{-1}(v_0)$ assures $bit$ in $P$. We conclude because we have $|f^{-1}(v)| \leq k$.
\end{proof}

Finally, by using the appropriate utility function and controlling
the lower bit of the leader election, we conclude the main result of
this section.

\RestatableSimulatedTreeGraphLeaderElectionImpossibility*
\begin{proof}
    Assume $n$ is even for simplicity. By Lemma \ref{SimulatedGraphCoinImpossibility}, there is no \kunbiased fair coin toss protocol for $G$ for every $\epsilon \leq \frac{1}{2}$. By \SectionText \ref{section:ProblemsRelations}, an \kunbiased FLE protocol gives a $(\frac{1}{2}n)$-$k$-unbiased fair coin toss protocol, therefore there is no \kunbiased FLE protocol for every $\epsilon \leq \frac{\frac{1}{2}}{\frac{1}{2}n}=\frac{1}{n}$ for $G$, as required.
\end{proof}
Theorem \ref{TreesImpossibility:LeaderElectionImpossibility}
generalizes the previous result by Abraham et al.
\cite{abraham2013distributed}  which gives $k=\lceil\frac{1}{2}n \rceil$ for a
general network, because every graph is a $\lceil \frac{1}{2}n \rceil$-simulated tree.
\begin{claim}
	Every connected graph is a $\lceil \frac{1}{2}n \rceil$-simulated tree.
\end{claim}
\begin{proof}
Given a connected graph $G(V,E)$, we build a partition of its vertices,
$B_1, \ldots, B_L $,
into connected sets of size at most $\frac{1}{2}n$ inductively.

For the first set, $B_1$, we take a connected set of size $\lceil \frac{1}{2}n \rceil$.
For each of the following sets, $B_i$, we take a maximal connected set
out of the vertices left, $V \backslash \bigcup_{j < i} B_j$.
Let $G'=(\{B_1,\ldots, B_L\}, E')$ be the graph induced over $B_1, ..., B_L$.
It is connected because $G$ is connected.

Assume by contradiction that $G'$ contains a cycle,
then that cycle has at least three vertices, and therefore there exist two
adjacent vertices in $G'$: $B_i, B_j$ such that $i,j\ne 1$. W.l.o.g assume $i < j$.
From the maximality of $B_i$, in the construction of $B_i$ we could include $B_j$
in $B_i$. Contradiction.
\end{proof}

\mySection{Adaption of \PhaseAsyncLeadTitle to Non-Consecutive ids}
\label{appendix:NonConsecutive}

In \SectionText \ref{section:BetterAlgo} and in \SectionText
\ref{section:PhaseAsyncLeadFull} we assumed for simplicity that the
processors are located consecutively along the ring. I.e., we assumed
that $2$ is neighbor of $1$ and $3$, $3$ is neighbor of $2$ and $4$ etc.
We enhance the protocol by adding an indexing phase prior to its execution.
The $origin$ sends a counter with the value $1$. Upon receiving the counter, 
each processor increments the counter by $1$ and then forwards it. This
way each processor is assigned a number, and uses this number to decide when
to perform validation.

Next, we adapt the proof to this generalization. Change the definition of
$s(h)$ to be the event that $h$ sends a validation message as the round's
validator. Similarly, define $r(h)$ to be the event that $h-1$ sends a 
validation 
message when $h$ is the round's validator. In the proof, we rely only on 
two facts that utilize the ids continuity, however these facts also in
the new model. The first fact, is continuity of validators along every honest 
segment. I.e., for an honest segment $I_j=(h_1, h_2,..h_{l_j})$, if $h_1$ 
behaves like a validator after performing $r$ rounds, then $h_2$ behaves like a 
validator after performing $r+1$ rounds, etc. The second fact, is that every 
honest processor behaves like a validator exactly once.

\mySection{When the ids are not Known Ahead}
\label{appendix:AnonymousRing}

In our model, we assume that the set of $id$s is known to all the
processors prior to the execution of the protocol, however
originally in \cite{abraham2013distributed,afek2014distributed} the
$id$s are not given, but learned during the execution of the
protocol. Their protocol includes a preceding wake-up phase, where
processors exchange $id$s and agree upon a direction for the ring.
For the attacks presented in \SectionText \ref{section:Attacks}, this is
not an issue, because we can extend the attacks by defining the
adversarial deviation to execute the wake-up phase honestly.
However, the resilience proofs from \SectionText
\ref{section:GoodBounds} and from \SectionText \ref{section:BetterAlgo} do need this
assumption.

The essential problem is that adversarial processors might leverage the
wake-up phase (which we do not describe) in order to transfer information
quickly. The adversaries might cause some honest processors complete the
wake-up phase before the others, and then abuse the mechanism of the
wake-up phase in order to transfer information about the secret values
of those honest processors who completed the wake-up. We suspect that
the proofs can be extended to consider also the wake-up phase, however
it remains an open question.

Additionally to the essential problem described above, there is another
technical problem. Since we defined the domain of a rational utility $u$
to be $[n]\cup \{\FAIL\}$, the problem is not well-defined for unknown $id$s.
Even worse, it is not clear how to define the problem such that the $id$s
are unknown and there exists a resilient FLE protocol. For example,
consider the following natural definition. Assume the $id$s are
taken from a known large space $\Sigma$. Define a utility function
$u: \Sigma \cup \{\FAIL\} \to [0,1]$ to be rational if $ u_p(\FAIL)=0$.
Additionally, require resilience to hold for every set
of $id$s, $\Omega \subset \Sigma$. The following rational utility $u_0$
demonstrates that our definition is not useful. Define
$\forall x\in \Sigma: u_0(x) = \mathbbm{1}[x\notin \Omega]$.
For every FLE protocol $P$ we have $ E_P[u] = 0 $. But an
adversarial coalition can lie about their $id$s and obtain an
expected utility of $E_D[u] = \frac{k}{n}$. So for this definition
of the problem, there is no \kresilient FLE protocol for a
unidirectional ring for every $k > 1$.

A good solution to the technical problem of defining resilience,
would be to consider to settle for an unbiased protocol.
Define a protocol to be \kunbiased if for every set
of $id$s, for every adversarial deviation: $\forall j\in \Sigma:
Pr(outcome=j) \leq \frac{1}{n}+\epsilon$. We conjecture that the
protocols \RingOriginalProtocolSpaced and \PhaseAsyncLead are
\kunbiased for similar values of $\epsilon$ and $k$ as we proved
when the $id$s are unknown and under the definition above.

Out of the many challenges in proving these resilience conjectures, we
address one specific issue - the adversaries might cause every
honest segment to believe it contains an $origin$ processor. In the
wake up phase presented by
\cite{abraham2013distributed,afek2014distributed}, the processors
exchange $id$s, and the processor with the lowest $id$ is selected
to be the $origin$. There cannot be two $origin$ processors in the
same honest segment because the set of $id$s perceived by every two
processors in the same honest segment is identical. However, an adversarial
coalition can cause an allocation of an $origin$ in every honest segment.
If the name space $\Sigma$ is large enough, then the adversaries can
do it by masking (setting to $0$) the higher bits
of the $id$ of every honest processor $h\in I_j$ when sending it to
other segments $I_i, i\ne j$. In order to recover the actual $id$s,
adversaries can encode the lost bits in their own $id$s.

In order to cope with the allocation of an $origin$ in every honest
segment, we can still assume all $origin$ are adversaries and then
translate \kresilience under this assumption to
$\frac{1}{2}k$-$\epsilon$-resilience. Asymptotically, the result is
equivalent.

	\newpage
	\bibliographystyle{plain}

\end{document}